\documentclass[12pt,a4paper]{article}

\usepackage{lineno}  

\usepackage{graphicx}  

\usepackage{xspace}
\usepackage{color}
\usepackage{colortbl}
\usepackage{subfigure}
\usepackage{amsmath}

\usepackage{ifthen} 
\newboolean{pdflatex}
\setboolean{pdflatex}{true} 
\usepackage{rotating} 
\newboolean{articletitles}
\setboolean{articletitles}{true} 

\newboolean{uprightparticles}
\setboolean{uprightparticles}{false} 
\usepackage{amssymb}
\usepackage{amsfonts}
\usepackage{upgreek}
\usepackage{lhcb-symbols-def}
\usepackage{hyperref}    
\usepackage[all]{hypcap} 

\usepackage{bm}
\usepackage{afterpage}

\newcommand{\bea}{\begin{eqnarray}}
\newcommand{\eea}{\end{eqnarray}}
\newcommand{\beq}{\begin{equation}}
\newcommand{\eeq}{\end{equation}}

\usepackage{cite}
\usepackage{mciteplus}

\begin{document}
\renewcommand{\thefootnote}{\fnsymbol{footnote}}
\setcounter{footnote}{1}



\begin{titlepage}

\belowpdfbookmark{Title page}{title}

\pagenumbering{roman}
\vspace*{-1.5cm}
\centerline{\large EUROPEAN ORGANIZATION FOR NUCLEAR RESEARCH (CERN)}
\vspace*{1.5cm}
\hspace*{-5mm}\begin{tabular*}{16cm}{lc@{\extracolsep{\fill}}r}
\vspace*{-12mm}\mbox{\!\!\!\includegraphics[width=.12\textwidth]{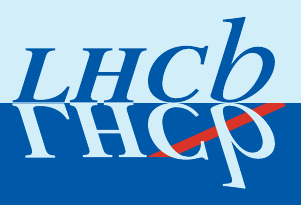}}& & \\
 & & CERN-PH-EP-2013-008\\
 & & LHCb-PAPER-2012-040\\  
 & & February 5, 2013 \\ 
 & & \\
\end{tabular*}

\vspace*{2.0cm}

{\bf\boldmath\Large
\begin{center}
Amplitude analysis and branching fraction measurement of $\Bsb\to \jpsi \Kp\Km$\\
\end{center}
}

\vspace*{1.0cm}
\begin{center}
\normalsize {
The LHCb collaboration\footnote{Authors are listed on the following pages.}
}
\end{center}

\begin{abstract}
  \noindent

An amplitude analysis of the final state structure in the  $\Bsb\to J/\psi K^+K^-$  decay mode is performed using $1.0~\rm fb^{-1}$ of data collected by the \lhcb experiment in 7~TeV center-of-mass energy $pp$ collisions produced by the LHC. A modified Dalitz plot analysis of the final state is performed  using both the invariant mass spectra and the decay angular distributions.  Resonant structures are observed in the $K^+K^-$ mass spectrum as well as a significant non-resonant S-wave contribution over the entire $K^+K^-$ mass range. The largest resonant component is the $\phi(1020)$, accompanied by $f_0(980)$, $f_2'(1525)$, and four additional resonances. The overall branching fraction is measured to be $\mathcal{B}(\Bsb \to J/\psi \Kp\Km)=(7.70\pm0.08\pm 0.39\pm 0.60)\times 10^{-4}$, where the first uncertainty is statistical, the second systematic, and the third due to the ratio of the number of \Bsb to $B^-$ mesons produced. The mass and width of the $ f_2'(1525)$ are measured to be $1522.2\pm 2.8^{+5.3}_{-2.0}~\rm MeV$ and $84\pm 6^{+10}_{-~5}~\rm MeV$, respectively. The final state fractions of the other resonant states are also reported.

\end{abstract}

\vspace*{2.0cm}
\vspace{\fill}


\vspace*{1.0cm}
\begin{center}
\hspace*{6mm}Submitted to Physical Review D\\
\end{center}
\vspace{\fill}

{\footnotesize
\centerline{\copyright~CERN on behalf of the \lhcb collaboration, license \href{http://creativecommons.org/licenses/by/3.0/}{CC-BY-3.0}.}}
\vspace*{2mm}

\clearpage
\newpage
\centerline{\large\bf LHCb collaboration}
\begin{flushleft}
\small
R.~Aaij$^{38}$, 
C.~Abellan~Beteta$^{33,n}$, 
A.~Adametz$^{11}$, 
B.~Adeva$^{34}$, 
M.~Adinolfi$^{43}$, 
C.~Adrover$^{6}$, 
A.~Affolder$^{49}$, 
Z.~Ajaltouni$^{5}$, 
J.~Albrecht$^{35}$, 
F.~Alessio$^{35}$
% ...

{\footnotesize \it
% ...
$ ^{5}$Clermont Universit\'{e}, Universit\'{e} Blaise Pascal, CNRS/IN2P3, LPC, Clermont-Ferrand, France\\
$ ^{6}$CPPM, Aix-Marseille Universit\'{e}, CNRS/IN2P3, Marseille, France\\
% ...
$ ^{11}$Physikalisches Institut, Ruprecht-Karls-Universit\"{a}t Heidelberg, Heidelberg, Germany\\
% ...
$ ^{33}$Universitat de Barcelona, Barcelona, Spain\\
$ ^{34}$Universidad de Santiago de Compostela, Santiago de Compostela, Spain\\
$ ^{35}$European Organization for Nuclear Research (CERN), Geneva, Switzerland\\
% ...
$ ^{38}$Nikhef National Institute for Subatomic Physics, Amsterdam, The Netherlands\\
% ...
$ ^{43}$H.H. Wills Physics Laboratory, University of Bristol, Bristol, United Kingdom\\
% ...
$ ^{49}$Oliver Lodge Laboratory, University of Liverpool, Liverpool, United Kingdom\\
% ...
$ ^{n}$LIFAELS, La Salle, Universitat Ramon Llull, Barcelona, Spain\\
% ...
}
\end{flushleft}






\end{titlepage}
\pagestyle{plain} 
\setcounter{page}{1}
\pagenumbering{arabic}



%
\clearpage
\section{Introduction}
\label{sec:Introduction}
\begin{figure}[hbt]
\centering
\includegraphics[scale=0.75]{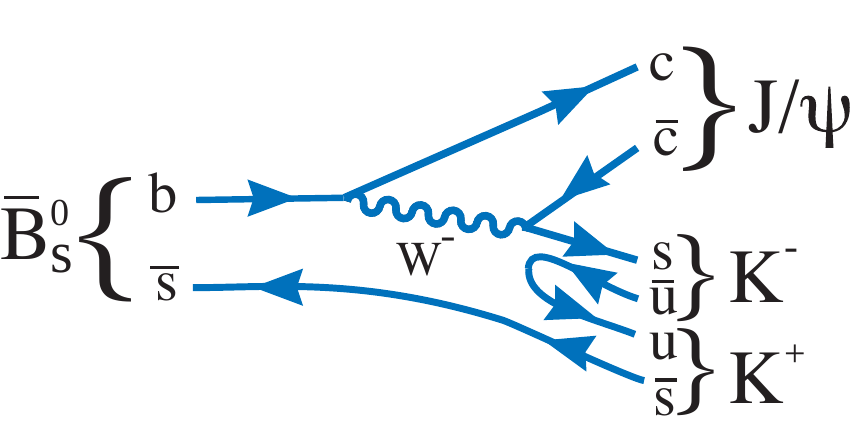}
\caption{\small Leading order diagram for $\Bsb\to J/\psi \Kp \Km$.}
\label{fig:BstoJpsiKK}
\end{figure}

The study of $\Bsb$ decays to $J/\psi h^+ h^-$, where $h$ is either a pion or kaon, has been used to measure mixing-induced \CP violation in $\Bsb$ decays \cite{LHCb:2012ad,LHCb:2011aa,LHCb-CONF-2012-002,CDF:2011af, Abazov:2011ry, Aaltonen:2012ie, :2012fu}.\footnote{Mention of a particular mode implies use of its charge conjugate throughout this paper.}  In order to best exploit these decays a better understanding of the final state composition is necessary. This study has been reported for the $\Bsb\to\jpsi\pi^+\pi^-$ channel \cite{LHCb:2012ae}. Here we perform a similar analysis for $\Bsb\to\jpsi K^+ K^-$.
While a large $\phi(1020)$ contribution is well known \cite{PDG} and the $f_2'(1525)$ component has been recently observed \cite{Aaij:2011ac} and confirmed \cite{Abazov:2012dz}, other components have not heretofore been identified including the source of  S-wave contributions~\cite{Stone:2008ak}.  The  tree-level Feynman diagram for the process is shown in Fig.  \ref{fig:BstoJpsiKK}.

In this paper the $J/\psi K^+$ and $K^+K^-$ mass spectra and decay angular distributions are used to study resonant and non-resonant structures. This differs from a classical ``Dalitz plot" analysis~\cite{Dalitz} since the $J/\psi$ meson has spin-1, and its three helicity amplitudes must be considered.  
\section{Data sample and detector}
\label{sec:Data}
The event sample is obtained using $1.0~\rm fb^{-1}$ of integrated luminosity collected with the \lhcb detector~\cite{LHCb-det} using $pp$ collisions at a center-of-mass energy of 7 TeV. The detector is a single-arm forward
spectrometer covering the pseudorapidity range $2<\eta <5$, designed
for the study of particles containing \bquark or \cquark quarks. Components include a high precision tracking system consisting of a
silicon-strip vertex detector surrounding the $pp$ interaction region,
a large-area silicon-strip detector located upstream of a dipole
magnet with a bending power of about $4{\rm\,Tm}$, and three stations
of silicon-strip detectors and straw drift-tubes placed
downstream.  The combined tracking system has momentum\footnote{We work in units where $c=1$.} resolution
$\Delta p/p$ that varies from 0.4\% at 5\gev to 0.6\% at 100\gev.  The impact parameter (IP) is defined as the minimum distance of approach of the track with respect to the primary vertex. For tracks with large
transverse momentum with respect to the proton beam direction, the IP resolution is  approximately 20\mum.  Charged hadrons are identified using two
ring-imaging Cherenkov detectors. Photon, electron and hadron
candidates are identified by a calorimeter system consisting of
scintillating-pad and preshower detectors, an electromagnetic
calorimeter and a hadronic calorimeter. Muons are identified by a 
system composed of alternating layers of iron and multiwire
proportional chambers. 

The trigger \cite{Aaij:2012me} consists of a hardware stage, based
on information from the calorimeter and muon systems, followed by a
software stage that applies a full event reconstruction. Events selected for this analysis are triggered by a $J/\psi\to\mu^+\mu^-$ decay, where the   
$J/\psi$ is required at the software level to be consistent with coming from the decay of a $\Bsb$ meson by use either of IP requirements or detachment of the $J/\psi$ from the primary vertex.
Monte Carlo simulations are performed
using \pythia~\cite{Sjostrand:2006za} with the specific tuning given in Ref.~\cite{LHCb-PROC-2011-005}, and the \lhcb detector description based on \geant~\cite{Agostinelli:2002hh,*Allison:2006ve} described in Ref.~\cite{LHCb-PROC-2011-006}.  Decays of $B$ mesons are based on \evtgen \cite{Lange:2001uf}.

\section{Signal selection and backgrounds}
\label{sec:2}
We select $\Bsb\to J/\psi K^+K^-$ candidates trying to simultaneously maximize the signal yield and reduce the background. Candidate $J/\psi\to\mu^+\mu^-$ decays are combined with a pair of kaon candidates of opposite charge, and then requiring that all four tracks are consistent with coming from a common decay point.
To be considered
a $J/\psi\to\mu^+\mu^-$ candidate, particles identified as muons of opposite charge are required to have  transverse momentum, $p_{\rm T}$, greater than 500\,MeV,  and 
form a vertex with fit $\chi^2$ per number of degrees of freedom (ndf) less than 11.  These requirements  give rise to a large $\jpsi$ signal over a small background~\cite{Aaij:2011fx}. Only candidates with a dimuon invariant mass between $-$48~MeV to +43 MeV relative to the observed $J/\psi$ mass peak are selected. The asymmetric requirement is due to final-state electromagnetic  radiation. The two muons are subsequently kinematically constrained to the known $J/\psi$ mass~\cite{PDG}. 

Our ring-imaging Cherenkov system allows for the possibility of positively identifying kaon candidates. Charged tracks produce Cherenkov photons whose emission angles are compared with those expected for electrons, pions, kaons or protons, and a likelihood for each species is then computed. To identify a particular species, the difference between the logarithm of the likelihoods for two particle hypotheses (DLL) is computed.  There are two criteria used: loose corresponds to DLL$(K-\pi)>0$, while tight has DLL$(K-\pi)>10$ and DLL$(K-p)>-3$. Unless stated otherwise, we require the tight criterion  for kaon selection.

We select candidate $\KpKm$ combinations if each particle is inconsistent with having been produced at the primary vertex. For this test
we require that the $\chi^2$  formed by using the hypothesis that the IP is zero be
greater than 9 for each track.
Furthermore,  each kaon must have  $p_{\rm T}>250$ MeV and the scalar sum of the $p_{\rm T}$ of the kaon candidates must be greater than 900 MeV.
To select $\Bsb$ candidates we further require that the
two kaon candidates  form a vertex with $\chi^2< 10$, and that they form a candidate $\Bsb$ vertex with the $J/\psi$ where the vertex fit $\chi^2$/ndf $<5$. We require that this $\Bsb$ vertex be more than $1.5$\,mm from the primary vertex, and the angle between the $\Bsb$ momentum vector and  the vector  from the primary vertex  to the $\Bsb$ vertex must be less than 11.8~mrad.

\begin{figure}[hbt]
\centering
\includegraphics[scale=0.5]{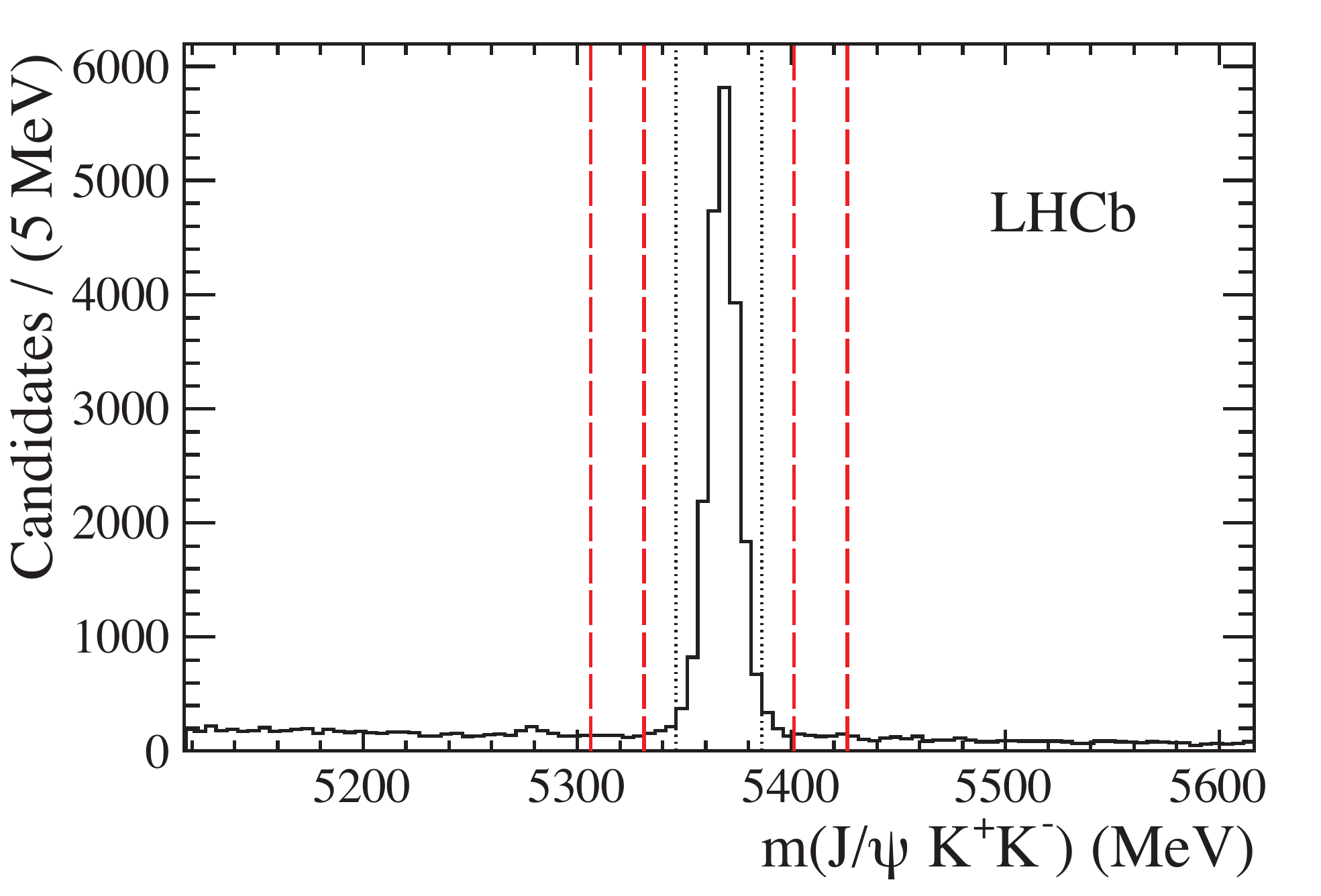}
\label{fig:Bs_mass}
\caption{\small Invariant mass spectrum of $J/\psi K^+K^-$ combinations. The vertical lines indicate the signal (black-dotted) and sideband (red-dashed) regions.}
\end{figure}

The $\Bsb$ candidate invariant mass distribution is shown in Fig.~\ref{fig:Bs_mass}. The vertical lines indicate the signal and sideband regions, where the signal region extends to $\pm 20$ MeV around the nominal $\Bsb$ mass \cite{PDG} and the sidebands extend from 35 MeV to 60 MeV on either side of the peak. The small peak near 5280 MeV results from $\Bdb$ decays, and will be subject to future investigation.

The background consist of combinations of tracks, which have a smooth mass shape through the $\jpsi\KpKm$  region, and peaking contributions caused by the reflection of specific decay modes where a pion is misidentified as a kaon. The reflection background that arises from the decay $\Bdb \to J/\psi K^-\pip$, where the $\pip$ is misidentified as a $K^+$, is determined from the number of \Bdb candidates in the control region $25-200$~MeV above the \Bsb mass peak.

\begin{figure}[hbt]
\centering
\includegraphics[scale=0.5]{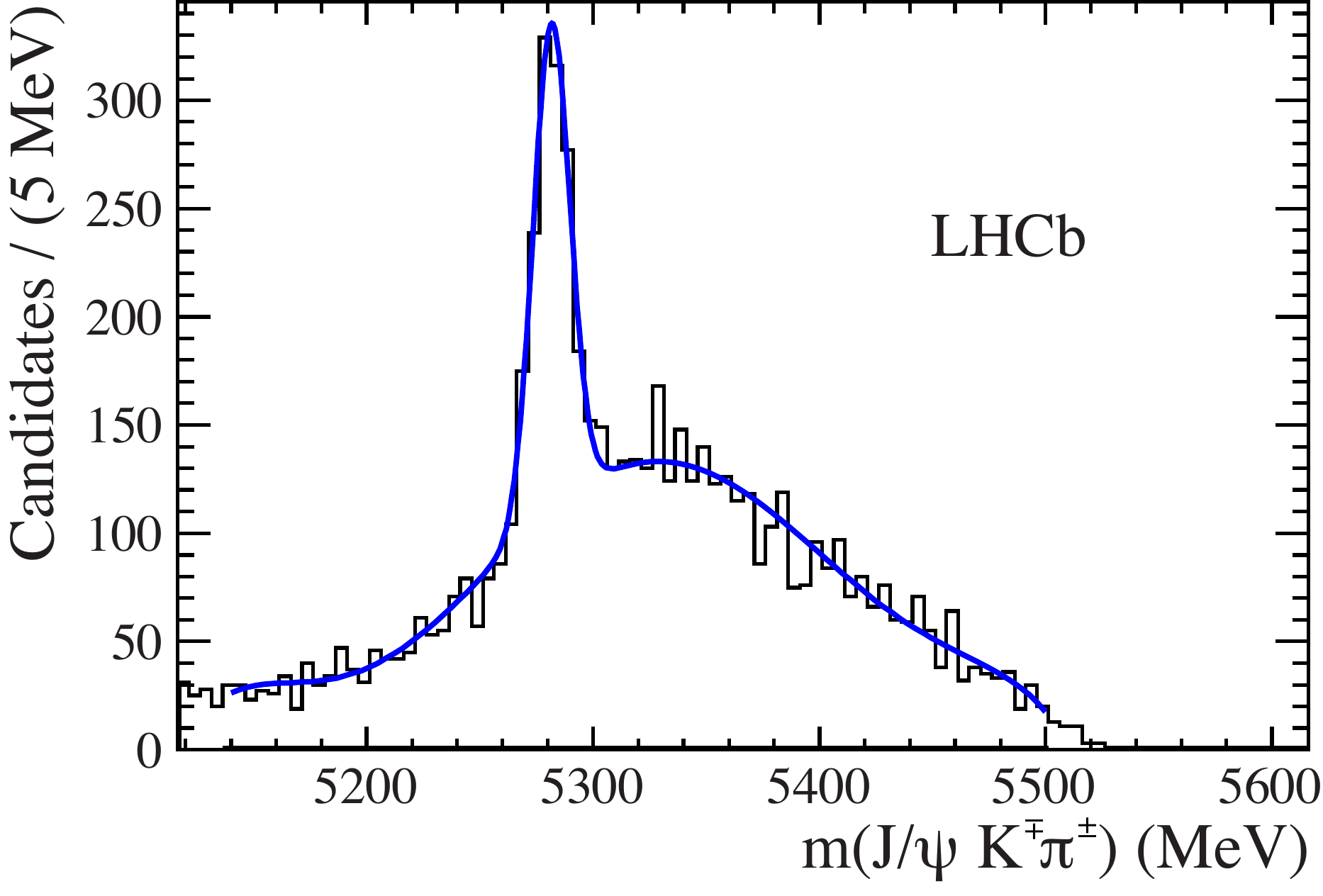}
\label{fig:reflection}
\caption{\small Invariant mass distribution for $J/\psi \Kp\Km$ candidates  $25-200$ MeV above the $\Bsb$ mass, reinterpreted as $\Bdb\to J/\psi K^{\mp} \pi^{\pm}$ events. The fit is to a signal Gaussian whose mass and width are allowed to vary as well as the polynomial background.}
\end{figure}

 For each of 
the candidates in the $J/\psi \Kp\Km$ control region, we reassign  each of the two kaons in turn to the pion mass hypothesis. The resulting $J/\psi K \pi$ invariant mass distribution is shown in Fig.~\ref{fig:reflection}.
The peak at the $\Bdb$ mass has $906\pm 51$ candidates, determined by fitting the data to a Gaussian  function for the signal, and a polynomial function for the background. From these events we estimate the number in the $\Bsb$ signal region, based on a  simulation of the shape of the reflected distribution as a function of $J/\psi K^-K^+$ mass. 
Using simulated  $\Bdb\to J/\psi \Kstarzb(892)$ and  $\Bdb\to J/\psi \Kstarb_2(1430)$ samples, we calculate $309 \pm 17$ reflection candidates within $\pm20$ MeV of the $\Bsb$ peak. This number is used as a constraint in the mass fit  described below.

\begin{figure}[h]
\centering
\includegraphics[scale=0.5]{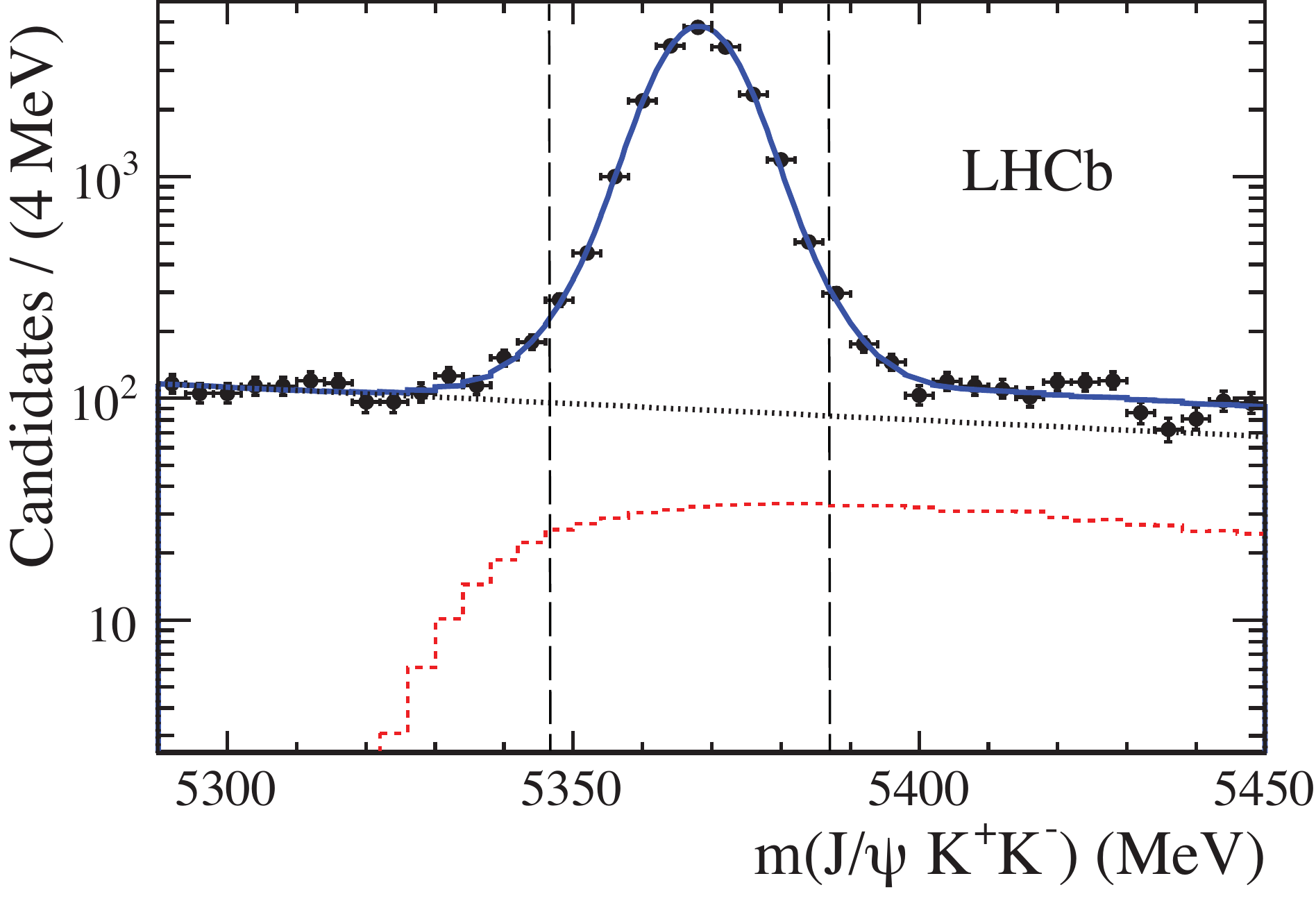}
\label{fig:Bs2JpsiKK}
\caption{\small Fit to the invariant mass spectrum of $J/\psi K^+K^-$ combinations.  The dotted (black) line is the combinatorial  background, the dashed (red) shape shows the misidentified  $\Bdb\to J/\psi \Km \pip$decays, and  the solid (blue) curve shows the total. The vertical dashed lines indicate the signal region.}
\end{figure}

To determine the number of \Bsb signal candidates we perform a fit to the candidate $J/\psi K^+K^-$ invariant mass spectrum shown in Fig.~\ref{fig:Bs2JpsiKK}.
The fit function is the sum of  the $\Bsb$ signal component,  combinatorial background, and the contribution from  the $\Bdb\to J/\psi \Km\pip$ reflections. The signal is modeled by a double-Gaussian function with a common mean. The combinatorial background is described by a linear function. The reflection background is constrained as described above. The mass fit gives 19,195$\pm$150 signal  together with $894 \pm 24$ combinatorial background candidates within $\pm 20$ MeV of the $\Bsb$ mass peak. 

\begin{figure}[ht]
\centering
\includegraphics[scale=0.5]{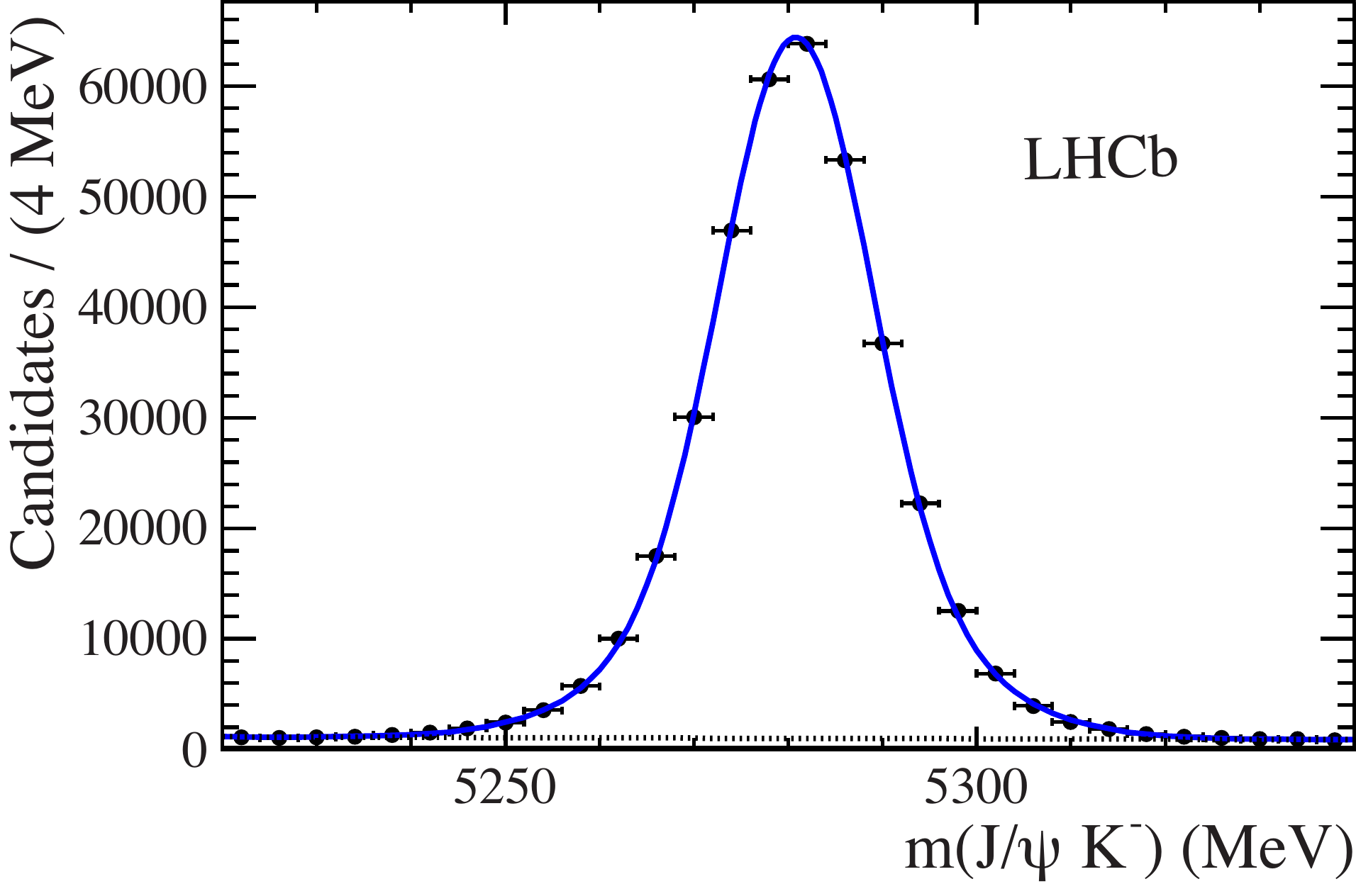}
\caption{\small Fit to the invariant mass spectrum of $J/\psi \Km$ candidates. The dotted line shows the combinatorial background and the solid (blue) curve is the total.}
\label{fig:Bu2JpsiK}
\end{figure}

We use the decay $\Bm\to \jpsi \Km$ as the normalization channel for branching fraction determinations. The selection criteria are similar to those used for $\jpsi\KpKm$, except for particle identification as here a loose kaon identification criterion  is used.  Figure~\ref{fig:Bu2JpsiK} shows the $J/\psi \Km$ mass distribution. The signal is fit with a  double-Gaussian function 
and a linear function is used to fit the combinatorial background. There are  342,786$\pm$661 signal and 10,195$\pm$134 background candidates  within $\pm20$ MeV of the $\Bm$ peak.


\section{Analysis formalism}
\label{sec:Formalism}
One of the goals of this analysis is to determine the intermediate states in $\Bsb\to J/\psi K^+K^-$ decay within the context of an isobar model~\cite{Fleming:1964zz, Morgan:1968zza}, where we  sum the resonant and non-resonant components testing if they explain the invariant mass squared and angular distributions.  We also determine the absolute branching fractions of $\Bsb\to\jpsi\phi(1020)$ and $\Bsb\to\jpsi f_2'(1525)$ final states and the mass and width of the $f_2'(1525)$ resonance. Another important goal is to understand the S-wave content in the $\phi(1020)$ mass region.

Four variables completely describe the decay of $\Bsb\to J/\psi K^+K^-$ with $J/\psi\to \mu^+\mu^-$. Two are the invariant mass squared of $J/\psi K^+$, $s_{12}\equiv m^2(J/\psi K^+)$, and the invariant mass squared of $K^+K^-$, $s_{23}\equiv m^2(K^+K^-)$. The other two are the $J/\psi$ helicity angle, $\theta_{J/\psi}$, which is the angle of the $\mu^+$ in the $J/\psi$ rest frame with respect to the $J/\psi$ direction in the  $\Bsb$ rest frame, and the angle between the $J/\psi$ and $K^+K^-$ decay planes, $\chi$, in the $\Bsb$ rest frame. 
  To simplify the probability density function (PDF), we analyze the decay process after integrating over the angular variable $\chi$, which eliminates  several interference terms. 

\subsection{\boldmath The model for $\Bsb\to J/\psi K^+K^-$ }
 
In order to perform an amplitude analysis a PDF must be constructed that models correctly the dynamical and kinematic properties of the decay.
The PDF is separated into two components, one describing signal, $S$, and the other background, $B$.
The overall PDF given by the sum is 
\begin{eqnarray}
\label{eq:pdf}
F(s_{12}, s_{23}, \theta_{J/\psi})&=&\frac{1-f_{\rm com }-f_{\rm refl}}{{\cal{N}}_{\rm sig}}\varepsilon(s_{12}, s_{23}, \theta_{J/\psi}) S(s_{12}, s_{23}, \theta_{J/\psi})\\ 
&+& B(s_{12}, s_{23}, \theta_{J/\psi})
,\nonumber 
\end{eqnarray}
where $\varepsilon$ is the detection efficiency. The background is described by the sum of combinatorial background, $C$, and reflection, $R$, functions
\begin{equation}
\label{eq:background}
B(s_{12}, s_{23}, \theta_{J/\psi})= \frac{f_{\rm com}}{{\cal{N}}_{\rm com}} C(s_{12}, s_{23},  \theta_{J/\psi})+ \frac{f_{\rm refl}}{{\cal{N}}_{\rm refl}} R(s_{12}, s_{23}, \theta_{J/\psi}),
\end{equation}
where $f_{\rm com}$ and $f_{\rm refl}$ are the fractions of the combinatorial background and reflection, respectively, in the fitted region. The fractions $f_{\rm com}$ and $f_{\rm refl}$ obtained from the mass fit are fixed for the subsequent analysis.

The normalization factors are given by
\begin{eqnarray}
{\cal{N}}_{\rm sig}&=&\int \! \varepsilon(s_{12}, s_{23}, \theta_{J/\psi}) S(s_{12}, s_{23}, \theta_{J/\psi}) \, 
ds_{12}\,ds_{23\,}d \cos \theta_{J/\psi},\nonumber\\
{\cal{N}}_{\rm com}&=&\int \!C(s_{12}, s_{23}, \theta_{J/\psi}) \, 
ds_{12\,}ds_{23\,}d\cos \theta_{J/\psi},\\
{\cal{N}}_{\rm refl}&=&\int \!R(s_{12}, s_{23}, \theta_{J/\psi}) \, 
ds_{12}\,ds_{23\,}d\cos \theta_{J/\psi}\nonumber.
\end{eqnarray}
This formalism similar to that used by Belle in their analysis of $\Bdb \to \Km\pi^+\chi_{c1}$~\cite{Mizuk:2008me}, and later used  by LHCb for the analysis of $\Bsb\to J/\psi \pi^+\pi^-$~\cite{LHCb:2012ae}.

\begin{figure}[t]
\centering
\includegraphics[scale=0.52]{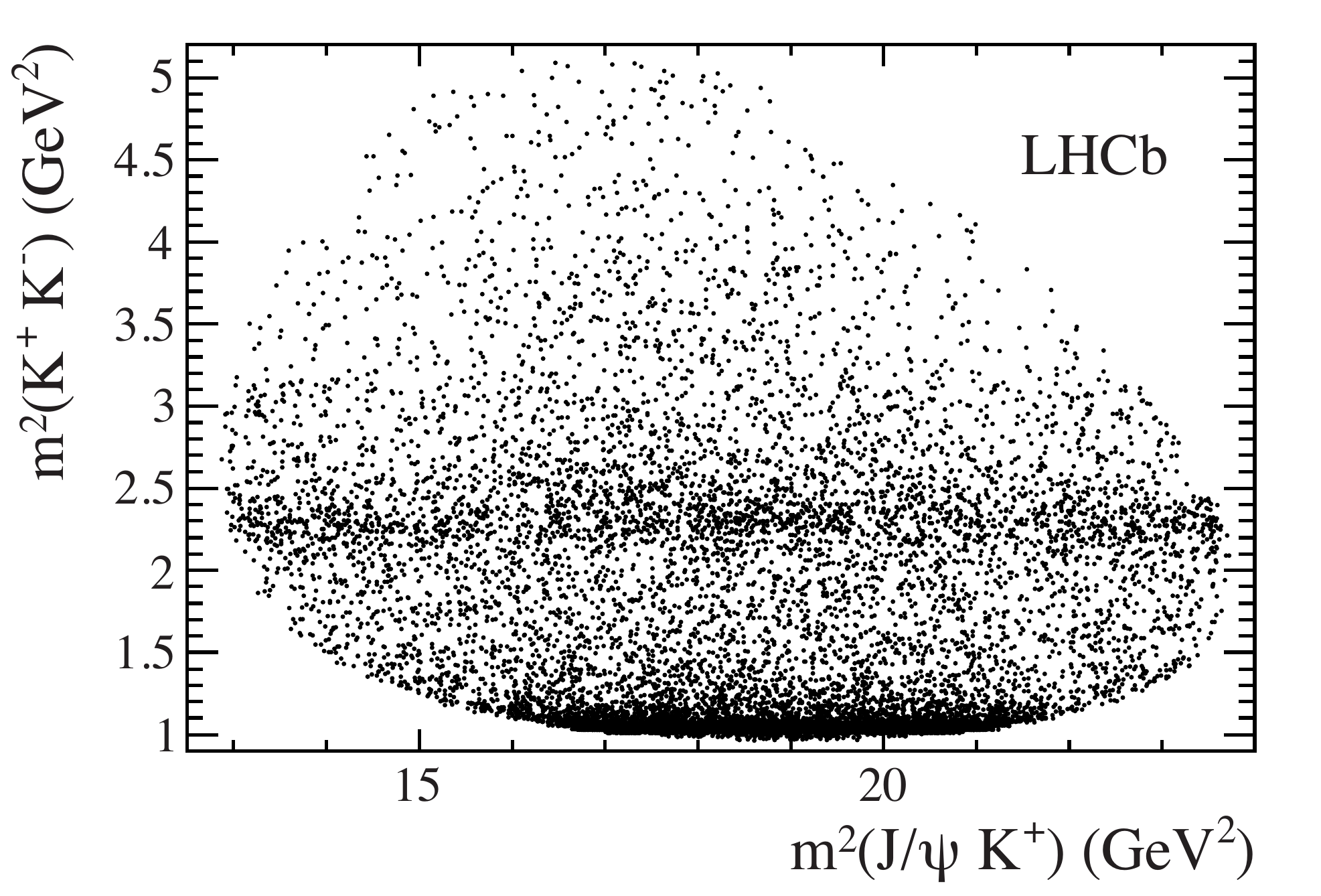}
\caption{\small Distribution of $m^2(K^+K^-)$ versus $m^2(J/\psi K^+)$ for $\Bsb$ candidate decays within $\pm 20$ MeV of the $\Bsb$ mass. The horizontal bands result from the $\phi(1020)$ and $f_2'(1525)$ resonances.}
\label{fig:dalitz}
\end{figure}

The invariant mass squared of $J/\psi K^+$ versus $K^+K^-$  is shown in Fig.~\ref{fig:dalitz} for $\Bsb\to J/\psi K^+K^-$ candidates.  No structure is seen in $m^2(\jpsi \Kp)$. There are however visible horizontal bands in the $K^+K^-$ mass squared spectrum, the most prominent of which correspond to the $\phi(1020)$ and $f_2'(1525)$ resonances. These and other structures in $m^2(K^+K^-)$ are now examined.

 
The signal function is given by the coherent sum over resonant states that decay into  $K^+K^-$, plus a possible non-resonant S-wave contribution\footnote{The interference terms between different helicities are zero because we integrate over the angular variable $\chi$.}
\begin{equation}
S(s_{12}, s_{23}, \theta_{J/\psi})=\sum_{\lambda=0,\pm1}\left|\sum_{i}a^{R_i}_{\lambda}e^{i\phi^{R_i}_{\lambda}}
\mathcal{A}_{\lambda}^{R_i}(s_{12}, s_{23}, \theta_{J/\psi})\right|^2, \label{amplitude-eq}
\end{equation} 
where $\mathcal{A}_{\lambda}^{R_i}(s_{12}, s_{23}, \theta_{J/\psi})$ describes the decay amplitude via an intermediate resonance state $R_i$ with helicity $\lambda$. Note that the $\jpsi$ has the same helicity as the intermediate $K^+K^-$ resonance. Each $R_i$ has an associated amplitude strength $a_{\lambda}^{R_i}$  and a phase $\phi_{\lambda}^{R_i}$ for each helicity state $\lambda$.  The amplitude for resonance $R$, for each $i$, is given by
\begin{equation}
\label{eq:dasAmp}
\mathcal{A}_{\lambda}^R(s_{12},s_{23}, \theta_{J/\psi})= F_B^{(L_B)}\; A_R(s_{23})\;F_R^{(L_R)}\; T_{\lambda} (\theta_{KK})
 \Big(\frac{P_B}{m_B}\Big )^{L_B}\; \Big( \frac{P_R}{\sqrt{s_{23}}}\Big )^{L_R}\; \Theta_{\lambda}(\theta_{J/\psi}),
\end{equation}
where  $P_R$ is the momentum of either of the two kaons in the di-kaon rest frame, $m_{B}$ is the $\Bsb$ mass, $P_B$  is the magnitude of the $J/\psi$ three-momentum in the $\Bsb$ rest frame, and $F_B^{(L_B)}$ and $F_R^{(L_R)}$ are the $\Bsb$ meson and $R_i$ resonance decay form factors.  The orbital angular momenta  between the $J/\psi$ and $K^+K^-$ system is given by  $L_B$, and the orbital angular momentum in the $K^+K^-$ decay is given by $L_R$; the latter is the same as the spin of the $K^+K^-$ system. Since the parent $\Bsb$ has spin-0 and the $J/\psi$ is a vector, when the $K^+K^-$ system forms a spin-0 resonance, $L_B=1$ and $L_R=0$. For $K^+K^-$ resonances with non-zero spin, $L_B$ can be 0, 1 or 2 (1, 2 or 3) for $L_R=1(2)$ and so on. We take the lowest $L_B$ as the default value and consider the other
possibilities in the systematic uncertainty.  

The Blatt-Weisskopf barrier factors $F_B^{(L_B)}$ and  $F_R^{(L_R)}$ \cite{Blatt}  are 
 \begin{eqnarray}
F^{(0)} &=& 1, \nonumber \\
F^{(1)} &=& \frac{\sqrt{1+z_0}}{\sqrt{1+z}},\\
F^{(2)} &=& \frac{\sqrt{z_0^2+3z_0+9}}{\sqrt{z^2+3z+9}}. \nonumber
\end{eqnarray}
For the $B$ meson  $z = r^2P_B^2$, where $r$, the hadron scale, is taken as 5.0 GeV$^{-1}$; for the $R$ resonance $z = r^2P_R^2$, and $r$ is taken as 1.5 GeV$^{-1}$ \cite{Kopp:2000gv}. In both cases $z_0= r^2P_0^2$ where $P_0$ is the decay  daughter momentum at the pole mass; for the \Bsb decay the \jpsi momentum is used, while for the $R$ resonances the kaon momentum is used.

In the helicity formalism, the angular term, $T_{\lambda}(\theta_{KK})$ is defined as 
\begin{equation}
 T_{\lambda}(\theta_{KK}) = d^J_{\lambda 0}(\theta_{KK}),
\end{equation}
where $d$ is the Wigner $d$-function,
 $J$ is the resonance spin, $\theta_{KK}$ is the helicity angle of the $K^+$ in the $K^+K^-$ rest frame with respect to the $\Kp\Km$ direction  in the $\Bsb$ rest frame, and may be calculated directly from the other variables as
\begin{equation}
\cos \theta_{KK} = \frac{\left[m^2(J/\psi K^+)-m^2(J/\psi K^-)\right]m(K^+K^-)}{4P_R P_B m_{B}}. \label{heli1}
\end{equation}
The $J/\psi$ helicity dependent term  $\Theta_{\lambda}(\theta_{J/\psi})$ is defined as
\begin{eqnarray}
 \Theta_{\lambda}(\theta_{J/\psi})&=& \sqrt{\sin^2\theta_{J/\psi}}\;\;\;\;\;\;\;\;\; \text{for}\;\; \lambda = 0 \nonumber \\
 &=&\sqrt{\frac{1+\cos^2\theta_{J/\psi}}{2}}\;\; \text{for}\;\; |\lambda| = 1. \label{heli2}
\end{eqnarray} 

The mass squared shape of each resonance, $R$ is described by the function $A_R(s_{23})$. 
In most cases this is a  Breit-Wigner (BW) amplitude. When a decay channel opens close to the resonant mass, complications arise, since the proximity of the second threshold distorts the line shape of the amplitude. The  $f_0(980)$ can decay to either $\pi\pi$ or $KK$. While the $\pi\pi$ channel opens at much lower masses, the $\Kp\Km$ decay channel opens near the resonance mass.  Thus, for the $f_0(980)$ we use a Flatt\'e model~\cite{Flatte:1976xv} that takes into account these coupled channels.
 
We describe the BW amplitude for a resonance decaying into two spin-0 particles, labeled as 2 and 3, as
\begin{equation}
A_R(s_{23})=\frac{1}{m^2_R-s_{23}-im_R\Gamma(s_{23})}~,
\end{equation}
where $m_R$ is the resonance mass, $\Gamma(s_{23})$ is its energy-dependent width that is parametrized as 
\begin{equation}
\Gamma(s_{23})=\Gamma_0\left(\frac{P_R}{P_{R_0}}\right)^{2L_R+1}\left(\frac{m_R}{\sqrt{s_{23}}}\right)F^2_R~.
\end{equation}
Here $\Gamma_0$ is the decay width when the invariant mass of the daughter combinations is equal to $m_R$.
 
The Flatt\'e mass shape is parametrized as 
\begin{equation}
A_R(s_{23})=\frac{1}{m_R^2-s_{23}-im_R(g_{\pi\pi}\rho_{\pi\pi}+g_{KK}\rho_{KK})},\label{Eq:Flatte}
\end{equation}
where the constants  $g_{\pi\pi}$ and $g_{KK}$ are the $f_0(980)$ couplings to $\pi^+\pi^-$ and $K^+K^-$ final states, respectively.
 The $\rho$ factors are given by Lorentz-invariant phase space
\begin{eqnarray}
\rho_{\pi\pi} &=& \frac{2}{3}\sqrt{1-\frac{4m^2_{\pi^{\pm}}}{s_{23}}}+\frac{1}{3}\sqrt{1-\frac{4m^2_{\pi^{0}}}{s_{23}}}\label{flatte1}, \\ 
\rho_{KK} &=& \frac{1}{2}\sqrt{1-\frac{4m^2_{K^{\pm}}}{s_{23}}}+\frac{1}{2}\sqrt{1-\frac{4m^2_{K^{0}}}{s_{23}}}.\label{flatte2}
\end{eqnarray}

For non-resonant processes, the amplitude $\mathcal{A}(s_{12},s_{23}, \theta_{J/\psi})$ is constant over the variables $s_{12}$ and $s_{23}$, but has an angular dependence due to the $J/\psi$ decay. The amplitude is derived from
Eq.~(\ref{eq:dasAmp}), assuming that the non-resonant $K^+K^-$ contribution is an S-wave (i.e. $L_R=0$, $L_B=1$) and is uniform in phase space (i.e. $A_R=1$),
\begin{equation}
\mathcal{A}(s_{12},s_{23}, \theta_{J/\psi}) = \frac{P_B}{m_B} \sqrt{\sin^2\theta_{J/\psi}}.
\end{equation}
\subsection{Detection efficiency}
\label{sec:efficiency}
The detection efficiency is determined from a phase space simulation sample containing $3.4\times 10^6$ $\Bsb\to J/\psi K^+K^-$   events with $J/\psi \to \mu^+\mu^-$. We also use a separate sample of $1.3\times 10^6$ $\Bsb\to J/\psi \phi$ events.  The $p$ and \pt distributions of the generated $\Bsb$ mesons are weighted to match the distributions found using $J/\psi\phi$ data.  The simulation is also corrected by weighting for difference between the simulated kaon detection efficiencies and the measured ones determined by using a sample of $D^{*+}\to \pip (D^0\to K^-\pip)$ events. 

Next we describe the efficiency in terms of the analysis variables. Both $s_{12}$ and $s_{13}$ range from 12.5 $\rm GeV^2$ to 24.0 $\rm GeV^2$, where $s_{13}$ is defined below, and thus are centered  at  $s_0=18.25~{\rm GeV}^2$. We model the detection efficiency using the 
dimensionless symmetric Dalitz plot observables
\begin{equation}
\label{eq:xandy}
x= \left(s_{12}-s_0\right)/\left(1~{\rm GeV}^2\right),~~~~~~  y=\left(s_{13}-s_0\right)/\left(1~{\rm GeV}^2\right),
\end{equation} 
and the angular variable $\theta_{J/\psi}$. The observables  $s_{12}$ and $s_{13}$ are related to $s_{23}$ as
\begin{equation}
s_{12}+s_{13}+s_{23}=m^2_B+m^2_{J/\psi}+m^2_{K^+}+m^2_{K^-}.\label{conver}
\end{equation}

\begin{figure}[htb]
\centering
\includegraphics[scale =0.5]{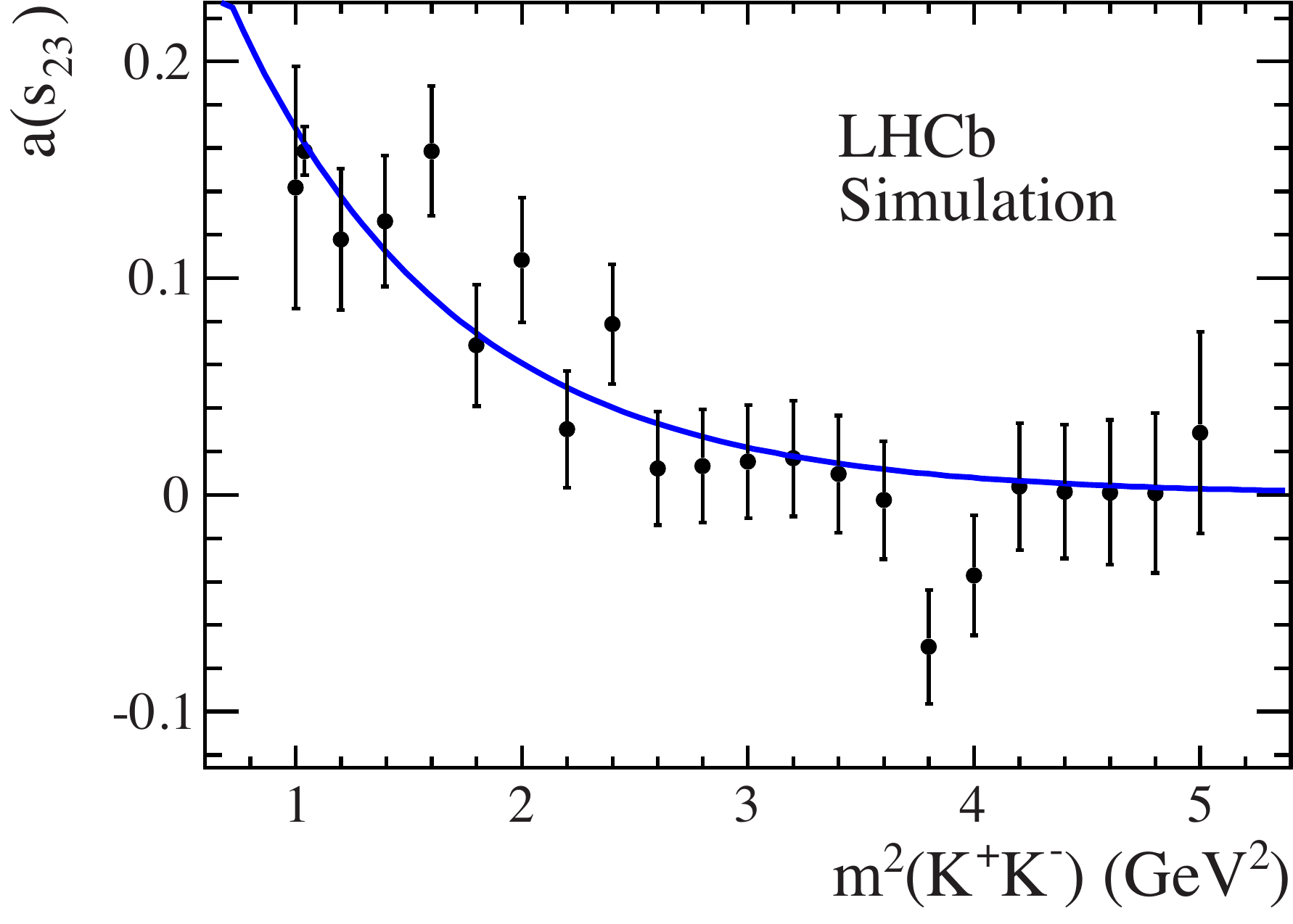}
\caption{\small Exponential fit to the efficiency parameter $a(s_{23})$. The point near the $\phi(1020)$ meson mass is determined more precisely due to the use of a large simulation sample.}
\label{fig:cosHacc}
\end{figure}

\begin{figure}[t]
\centering
\includegraphics[scale=0.43]{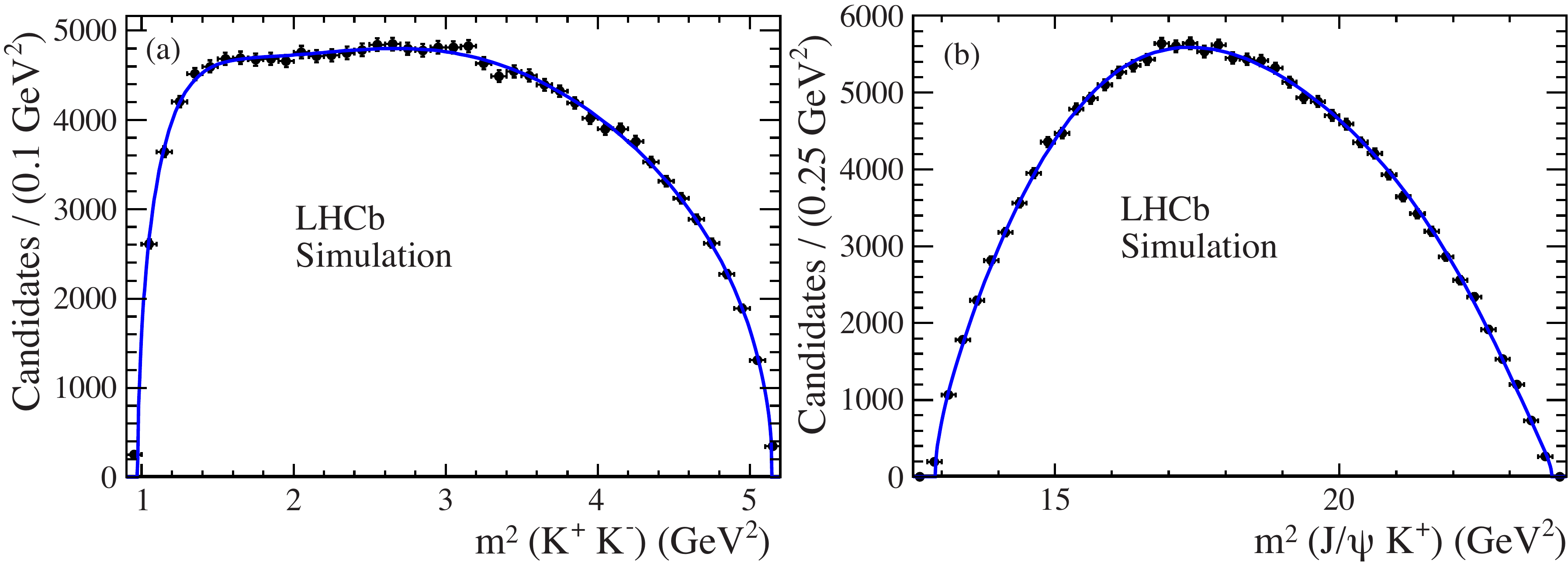}
\caption{\small Projections of  the invariant mass squared  (a) $K^+K^-$  and (b) $J/\psi K^{+}$ from the simulation used to measure the efficiency parameters. The points represent the generated event distributions and the curves the polynomial fit.}
\label{fig:effmodel2}
\end{figure}

To parametrize this efficiency, we fit the $\cos \theta_{J/\psi}$ distributions of  the $J/\psi K^+K^-$ and  $J/\psi \phi$  simulation samples in bins of $m^2(K^+K^-)$ with the function
\begin{equation}
\varepsilon_2(s_{23},\theta_{J/\psi})=\frac{1+ a\cos^2\theta_{J/\psi}}{2+2a/3},\label{eq:cosHacc}
\end{equation}
giving values of $a$ as a function of $m^2(K^+K^-)$. The resulting distribution, shown in Fig.~\ref{fig:cosHacc}, is  described by an exponential function
\begin{equation}
a(s_{23})= \exp(a_1+a_2 s_{23}),
\end{equation}
with $a_1= -0.76\pm 0.18$ and $a_2=(-1.02\pm 0.15)~\rm GeV^{-2}$. Equation~(\ref{eq:cosHacc}) is normalized with respect to $\cos \theta_{J/\psi}$. 
The efficiency in $\cos \theta_{J/\psi}$  depends on $s_{23}$, and is observed to be independent of $s_{12}$. 
Thus the detection efficiency  can be expressed as 
\begin{equation}
\varepsilon(s_{12}, s_{23}, \theta_{J/\psi})=\varepsilon_1(x,y)\times \varepsilon_2(s_{23}, \theta_{J/\psi}).\label{eq:eff}
\end{equation}
After integrating over $\cos \theta_{J/\psi}$,  Eq.~(\ref{eq:eff}) becomes
\begin{equation}
\int_{-1}^{+1}\varepsilon(s_{12}, s_{23}, \theta_{J/\psi})d\cos \theta_{J/\psi}=\varepsilon_1(x,y),
\end{equation}
and is modeled by 
 a symmetric fifth order polynomial function given by
\begin{eqnarray}
\varepsilon_1(x,y)&=& 1+\epsilon'_1(x+y)+\epsilon'_2(x+y)^2+\epsilon'_3xy+\epsilon'_4(x+y)^3
+\epsilon'_5 xy(x+y)\nonumber \\
&&+\epsilon'_6(x+y)^4+\epsilon'_7 xy(x+y)^2+\epsilon'_8 x^2y^2 +\epsilon'_9 (x+y)^5\nonumber\\
&&+\epsilon'_{10} xy(x+y)^3+\epsilon'_{11} x^2y^2(x+y),
\end{eqnarray}
where $\epsilon'_i$ are the fit parameters.
 \begin{figure}[b]
\centering
\includegraphics[scale=0.5]{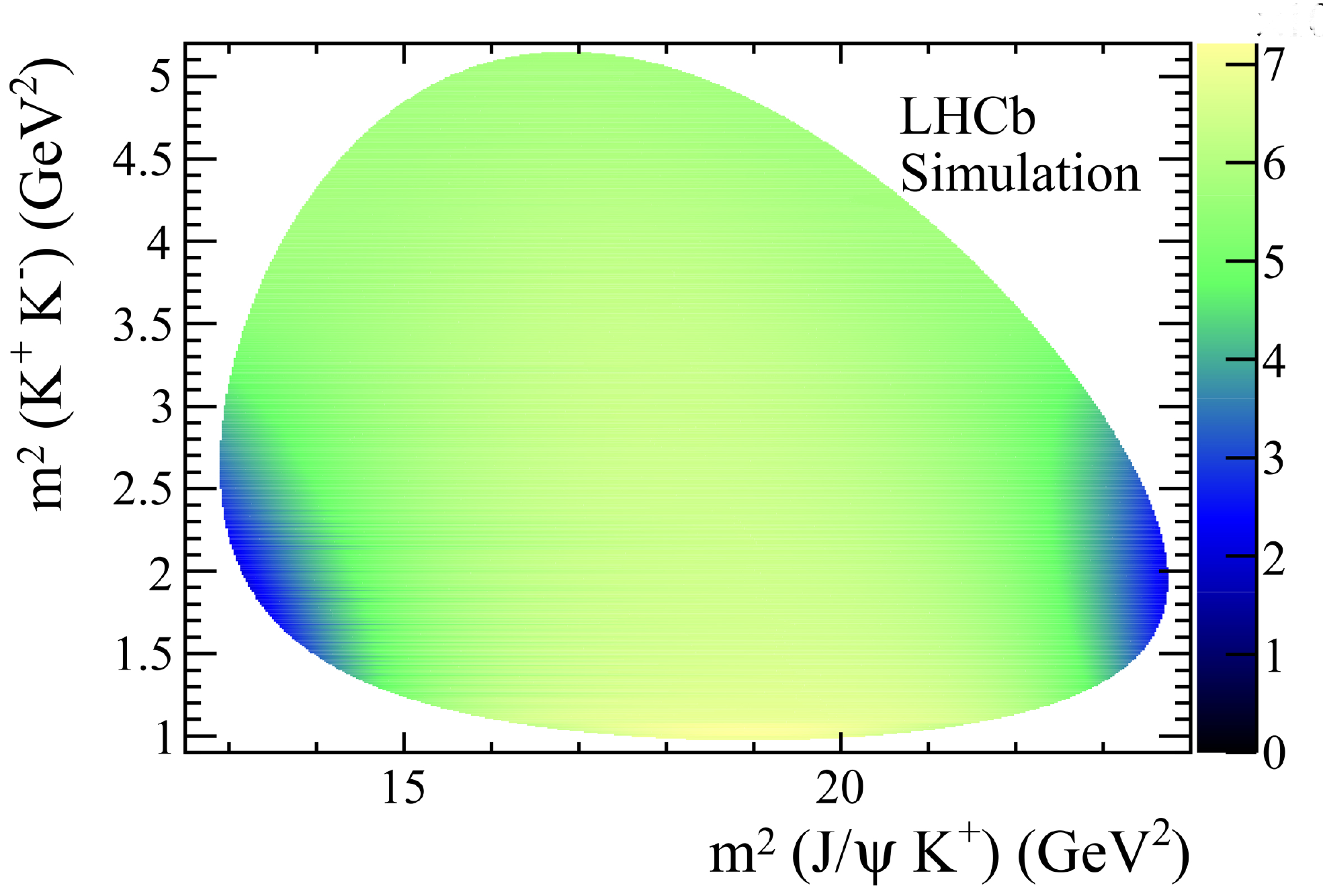}
\caption{\small Parametrized detection efficiency as a function of $m^2(K^+K^-)$ versus $m^2(J/\psi K^+)$. The z-axis scale is arbitrary.}
\label{fig:effmodel1}
\end{figure}
The $\Bsb \to J/\psi \Kp\Km$ phase space simulation sample is modeled with the polynomial function. The fitted  function is shown in Fig.~\ref{fig:effmodel1}, and the projections of the fit  are shown in Fig.~\ref{fig:effmodel2}. 
The  efficiency is well described by the parametrization. 

\begin{figure}[bh]
\centering
\includegraphics[scale =0.5]{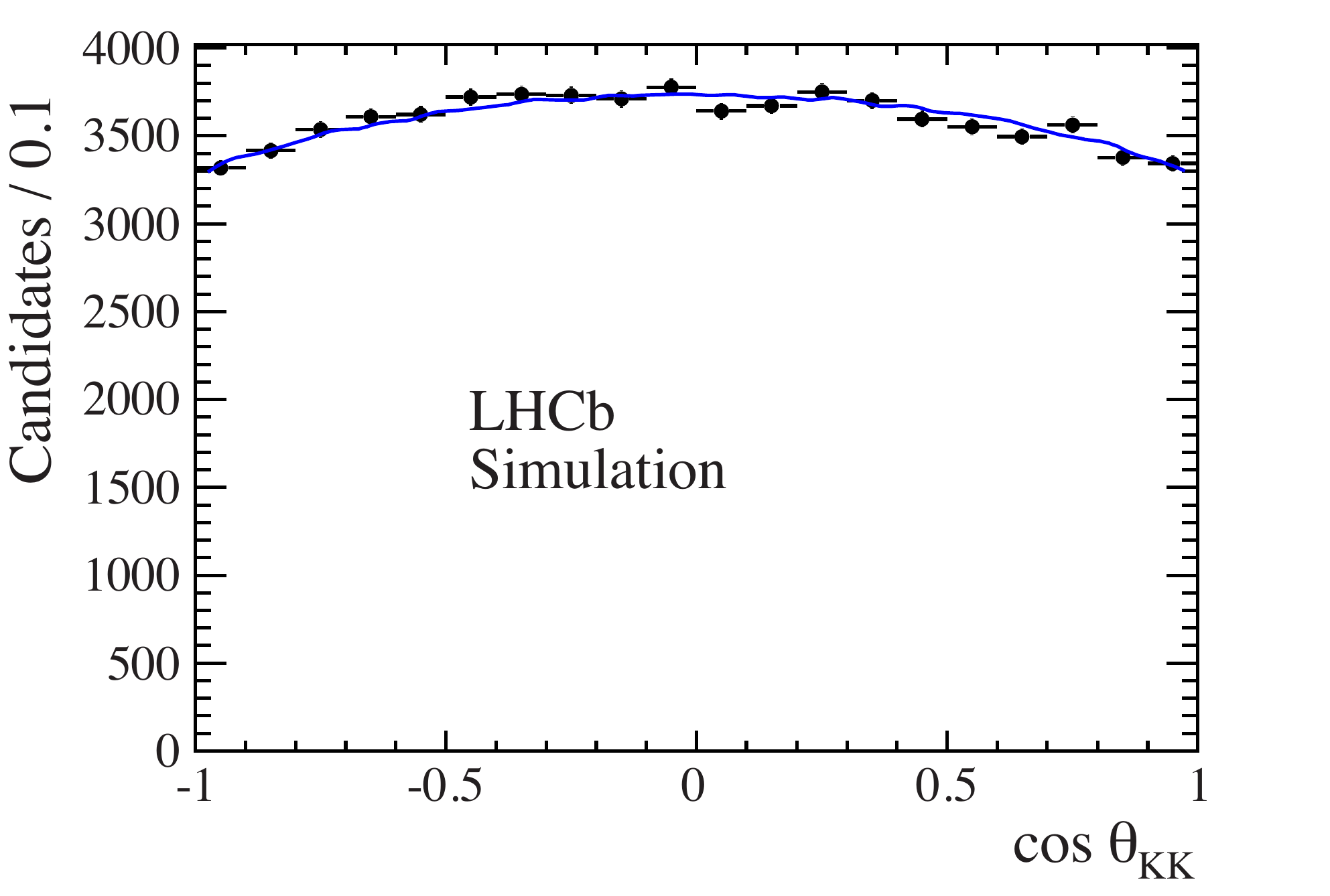}
\caption{\small Distribution of $\cos \theta_{KK}$  for the $J/\psi \phi$ simulated sample fitted with $\varepsilon_1(x,y
)\times A(\theta_{KK})$, within $\pm 20$ MeV of the $\phi(1020)$ mass.}
\label{fig:effcoskk}
\end{figure}

For the region within $\pm 20$ MeV of the $\phi(1020)$ mass, the $\cos \theta_{KK}$ acceptance is used separately, due to the large number of signal events. Here the $\cos \theta_{KK}$ distribution shows a variation in efficiency, which can be parametrized using the efficiency function 
\begin{equation}
A(\theta_{KK})=\frac{1+\epsilon'_{12} \cos^2\theta_{KK}}{1+\epsilon'_{12}/3},
\end{equation}
where the parameter $\epsilon'_{12}$ is measured from a  fit to the simulated $J/\psi \phi$ sample with $\varepsilon_1(x, y)\times A(\theta_{KK})$, giving $\epsilon'_{12}=-0.099\pm 0.010$, as shown in Fig.~\ref{fig:effcoskk}. 

The mass resolution is $\sim0.7$~MeV at the $\phi(1020)$ mass peak, which is added to the fit model by increasing the Breit-Wigner width of the $\phi(1020)$ to 4.59~MeV.

\subsection{Background composition}
\label{sec:background}

The  shape of the combinatorial background is modeled as
\begin{equation}
C(s_{12}, s_{23}, \theta_{J/\psi})=\left[C_1(s_{12}, s_{23})\frac{P_B}{m_B}+\frac{c_0}{(m^2_0-s_{23})^2+m^2_0\Gamma_0^2}\right]\times \left(1+\alpha \cos^2 \theta_{J/\psi}\right),
\end{equation}
where $C_1(s_{12}, s_{23})$ is  parametrized as
\begin{equation}
C_1(s_{12}, s_{23}) = 1+c_1(x+y)+c_2(x+y)^2+c_3xy+c_4(x+y)^3
+c_5xy(x+y),
\end{equation}
with $c_i$, $m_0$, $\Gamma_0$ and $\alpha$ as the fit parameters. The variables $x$ and $y$ are defined in Eq.~(\ref{eq:xandy}).

\begin{figure}[b]
\centering
\hspace*{-2mm}\includegraphics[scale=.41]{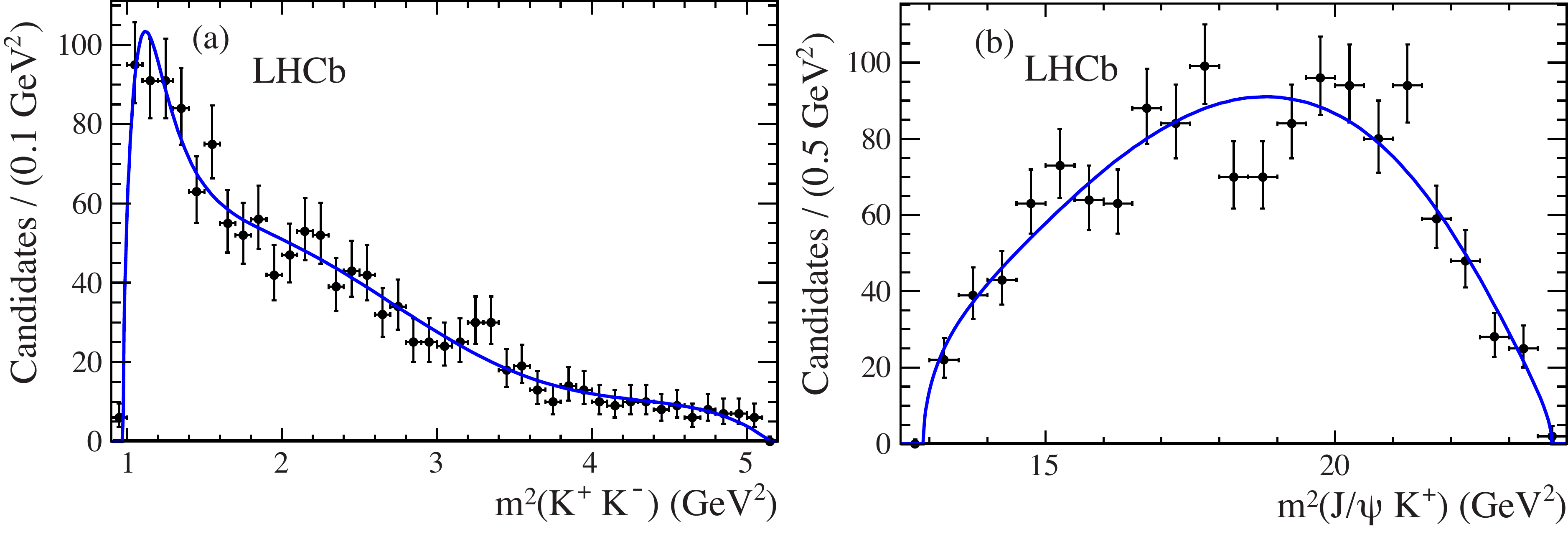}
\caption{\small Invariant mass squared projections of (a)  $K^+K^-$  and (b) $J/\psi K^{+}$ from the background Dalitz plot of candidates in the \Bsb mass sidebands.}
\label{fig:bkgmodel}
\end{figure}

\begin{figure}[t]
\centering
\includegraphics[scale=0.5]{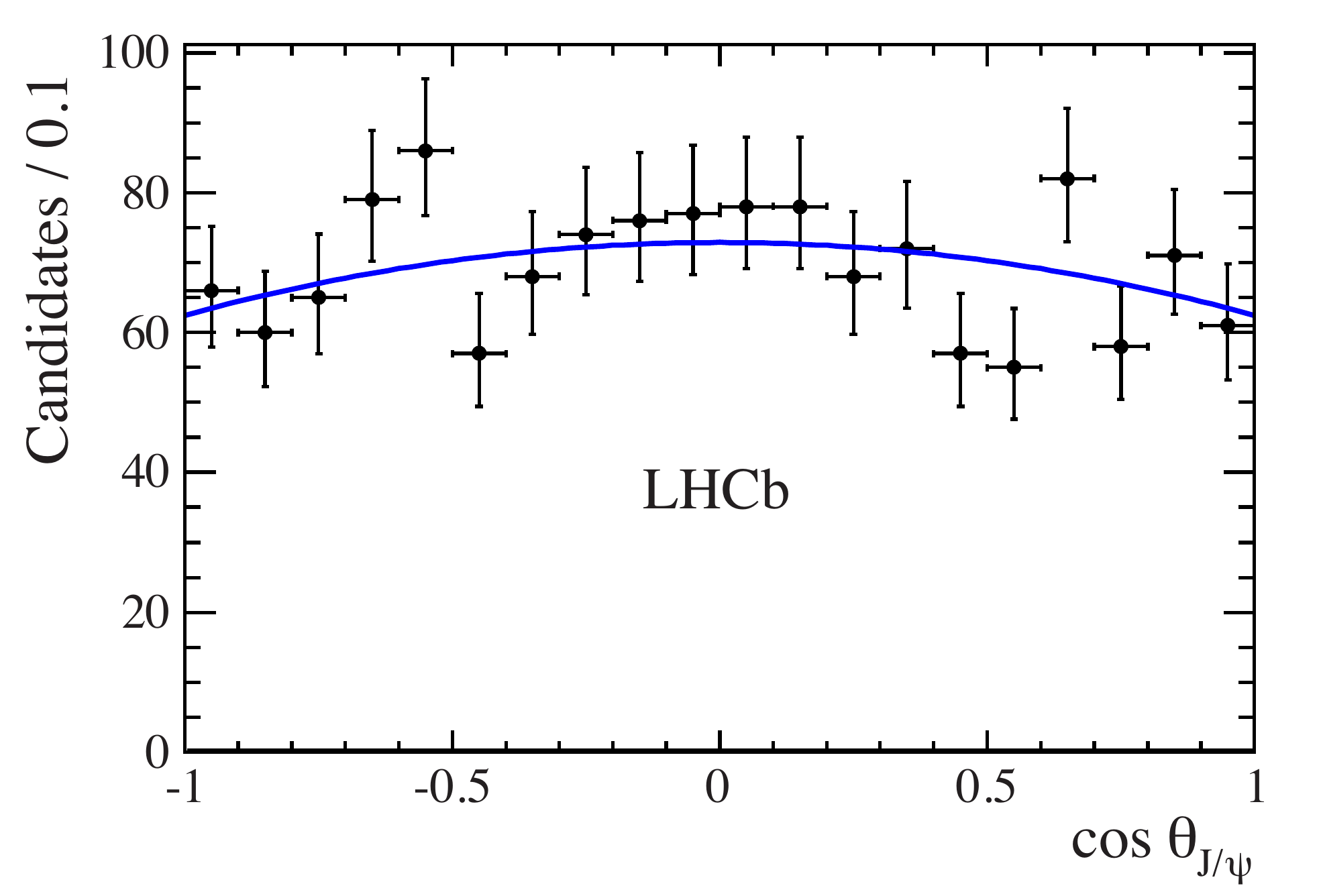}
\caption{\small Distribution of $\cos \theta_{J/\psi}$ from the background sample fit with the function $1+\alpha \cos^{2} \theta_{J/\psi}$.}
\label{fig:bkgcosH}
\end{figure}

 
Figure \ref{fig:bkgmodel} shows the mass squared projections from the $\Bsb$ mass sidebands with the fit projections overlaid. The $\chi^2/\rm ndf$ of the fit is 291/305.
The value of $\alpha$ is determined by fitting the $\cos\theta_{J/\psi}$ distribution of background, as shown in Fig. \ref{fig:bkgcosH}, with a function of the form $1+\alpha \cos^{2} \theta_{J/\psi}$, yielding $\alpha=-0.14\pm 0.08$.

\begin{figure}[htb]
\centering
\includegraphics[scale=0.42]{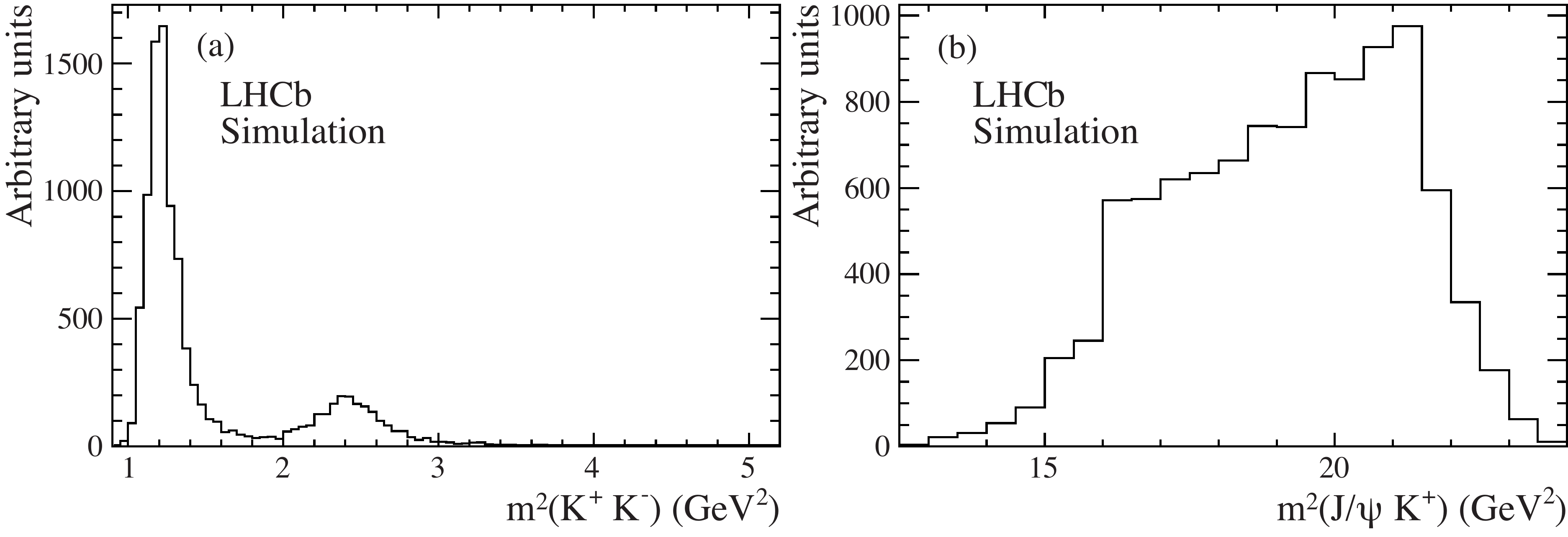}
\caption{\small Projections of the reflection background in the variables (a) $ m^2(K^+K^-)$   and (b) $ m^2(J/\psi K^{+})$, obtained from $\Bdb\to J/\psi \Kstarzb(892)$ and  $\Bdb\to J/\psi \Kstarb_2(1430)$ simulations.}
\label{fig:refmodel}
\end{figure}

The reflection background is parametrized as
\begin{equation}
R(s_{12}, s_{23}, \theta_{J/\psi})=R_1(s_{12}, s_{23})\times  \left(1+\beta \cos^2 \theta_{J/\psi}\right), 
\end{equation}
where $R_1(s_{12}, s_{23})$ is modeled using the simulation; the projections of $s_{12}$ and $s_{23}$ are shown in Fig.~\ref{fig:refmodel}.
\begin{figure}[t]
\centering
\includegraphics[scale=0.5]{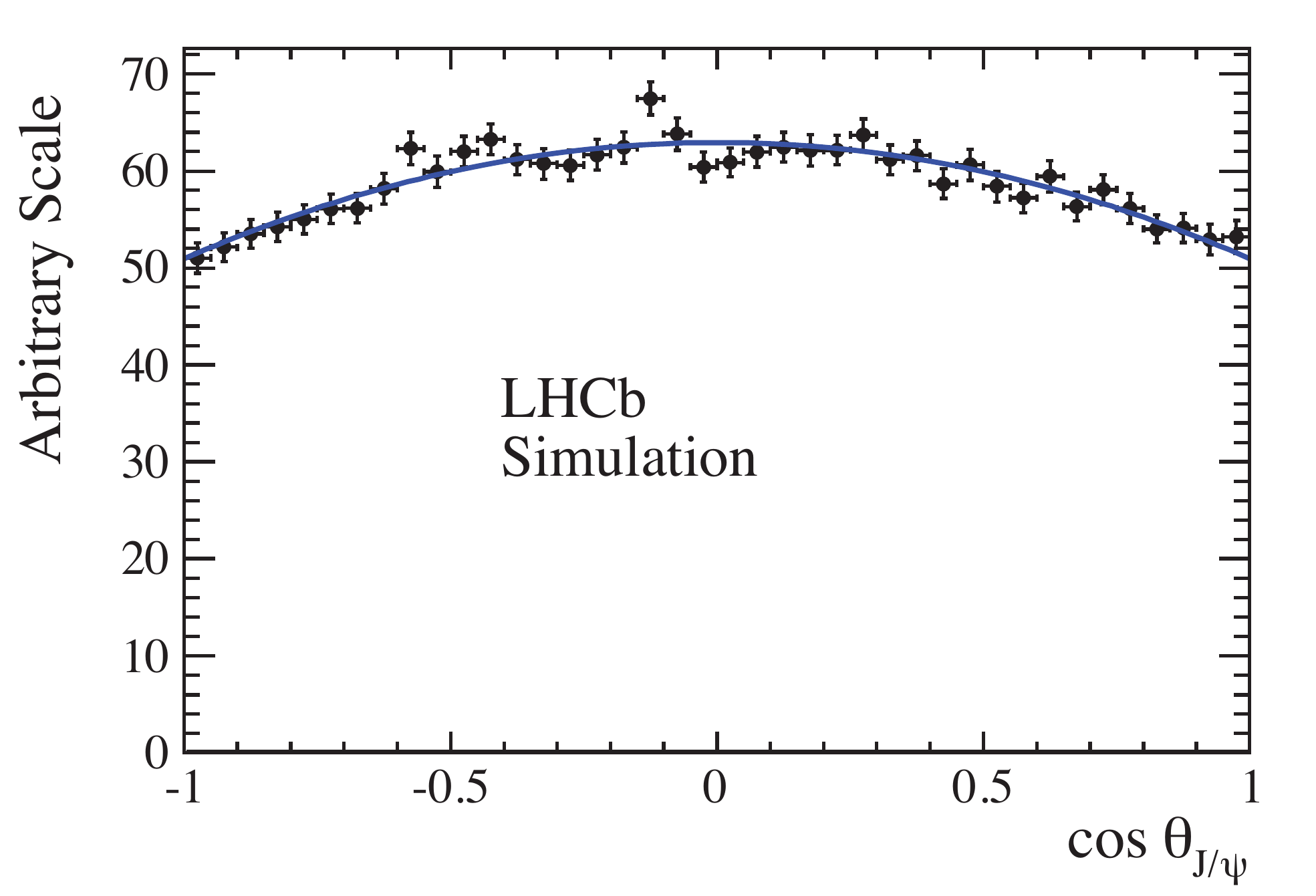}
\caption{\small Distribution of $\cos \theta_{J/\psi}$  for the reflection fit with the function $1+\beta \cos^{2} \theta_{J/\psi}$.}
\label{fig:refcosH}
\end{figure}
The $J/\psi$ helicity angle dependent part of the reflections is modeled as $1+\beta \cos^2 \theta_{J/\psi}$, where the parameter $\beta$ is obtained from a fit to the simulated $\cos \theta_{J/\psi}$ distribution, shown in Fig.~\ref{fig:refcosH}, giving $\beta= -0.19\pm 0.01$.

\section{Final state composition}
\subsection{Resonance models}
\label{sec:Resonance_models}
The resonances that are likely to contribute are produced from the  $s\bar{s}$ system in Fig.~\ref{fig:BstoJpsiKK}, and thus are isoscalar ($I=0$).
The $K^+K^-$ system in the decay  $\Bsb\to J/\psi K^+K^-$  can, in principle, have zero or any positive  integer angular momentum. Both the $P$-parity and $C$-parity of $\Kp \Km$ pair in a state of relative angular momentum $L$ are given by $(-1)^{L}$.  Therefore the allowed resonances decaying to $K^+K^-$ are limited to $J^{PC}=0^{++},~ 1^{--},~ 2^{++},~ ...,$ with isospin $I=0$. In the kinematically accessible mass range up to 2 GeV, resonances with $J^{PC}=3^{--}$ or higher are not expected and thus the subsequent analysis only uses spins up to $J=2$.  Possible resonance candidates  are listed in Table~\ref{tab:reso1}. There could also be a contribution from  non-resonant events which we assume to be S-wave and  evenly distributed over the available phase space.

\begin{table}[h]
\centering
\caption{Possible resonance candidates in the $\Bsb\to J/\psi K^+K^-$ decay mode.}
\vspace{0.2cm}
\begin{tabular}{cclcc}
\hline
Spin & Helicity & Resonance  & Amplitude \\
\hline
0 & 0 & $f_0(980)$ & Flatt\'e \\
0 & 0 & $f_0(1370)$, $f_0(1500)$, $f_0(1710)$ & BW\\
\hline
1 & $0,\pm1$ & $\phi(1020)$, $\phi(1680)$ & BW\\
\hline
2 & $0,\pm1$ & $f_2(1270)$, $f_2'(1525)$, $f_2(1640)$,& BW\\
&& $f_2(1750)$, $f_2(1950)$&\\
\hline
\end{tabular}\label{tab:reso1}
\end{table}

To study the resonant structures of the decay $\Bsb\to J/\psi K^+K^-$ we use 20,~\!425 candidates with an invariant mass within $\pm 20$ MeV of the observed \Bsb mass peak. This includes both signal and background, with  $94\%$ signal purity. 
We begin our analysis  considering only the resonance components $\phi(1020),~f_2'(1525)$ and a non-resonant component, established in our earlier measurement \cite{Aaij:2011ac}, and add  resonances until no others are found with more than two standard deviation statistical significance (2$\sigma$). The significance is estimated from the fit fraction divided by its statistical uncertainty. Our best fit model includes a non-resonant component and 8 resonance states:  $\phi(1020)$, $f_0(980)$, $f_0(1370)$, $f_2'(1525)$, $f_2(1640)~(\rm |\lambda|=1)$, $\phi(1680)~(\rm |\lambda|=1),$\footnote{The $f_2(1640)~(\lambda =0)$ and $\phi(1680)~(\lambda=0)$ components have less than two standard deviation significance when added separately to the fit, and therefore are not included in the best fit model.}  $f_2(1750)$, and $f_2(1950)$. Most of the resonances considered here are well established except for the modes $f_2(1640)$, $f_2(1750)$, and $f_2(1950)$. Although the existence of $f_2(1640)$ is not confirmed yet~\cite{PDG}, the right shoulder of $f_2'(1525)$ fits better when we add this state. The presence of multiple broad overlapping resonances in this region may indicate a failure of the isobar model used in this analysis, but with the present data sample alternative descriptions are not feasible.
Indeed, the situation is not clear for the resonance states in the vicinity of 1750 MeV. The PDG lists a spin-0 resonance, $f_0(1710)$, around 1.72 GeV of $\Kp\Km$ invariant mass~\cite{PDG}. The \belle collaboration observed  a  resonance in the vicinity of 1.75 GeV with $J^{PC}=(\rm even)^{++}$ in their study of $\gamma \gamma \to \Kp\Km$~\cite{Abe:2003vn}, but could not establish its spin.  A state of mass 1767$\pm$14 MeV was seen by the L3 collaboration decaying into $K^0_SK^0_S$ with $J=2$ \cite{Acciarri:2000ex}. 
We find that our data are better fit including the $f_2(1750)$ mode. If we substitute either the 
$f_0(1710)$ or $f_0(1750)$ resonance the fit is worsened, as the $\rm -ln\mathcal{L}$ increase by 59 and 7 units, respectively.

 In the same analysis of $\gamma \gamma \to \Kp\Km$, \belle also observed the $f_2(1950)$~\cite{Abe:2003vn}  resonance. We include this state in our best fit model. 
Furthermore, we do not expect significant contributions from the  $f_2(1270)$ and $f_0(1500)$ resonances, since the PDG branching fractions are much larger in the $\pip\pim$ final state than in $K^+K^-$~\cite{PDG} and we did not see significant contributions from these two resonances in the $\Bsb\to J/\psi\pip\pim$ final state~\cite{LHCb:2012ae}. Therefore, these two resonances are not considered in the best fit model. However, we add these states,  in turn, to the best fit model in order to test for their possible presence.

The masses and widths of the BW resonances are listed in Table~\ref{tab:resparam}. When used in the fit they are fixed to the central values, except for the $f_2'(1525)$, whose mass and width are allowed to vary. 

\begin{table}[h]
\centering
\caption{Breit-Wigner resonance parameters.}
\vspace{0.2cm}
\begin{tabular}{crclrcll}
\hline
 Resonance &\multicolumn{3}{c}{Mass (MeV)} & \multicolumn{3}{c}{\!\!\!\!Width (MeV)}& ~Source \\
 \hline
$\phi(1020)$ & $1019.46 $ \!\!\!\!\!\!\!&$\pm$& \!\!\!\!\!\!$ 0.02$ &$4.26$\!\!\!\!\!\!&$\pm$&\!\!\!\!\!\!$0.04$&PDG \cite{PDG}\\ 
$f_2(1270)$ & $1275$ \!\!\!\!\!\!\!\!&$\pm$& \!\!\!\!\!\!$ 1$ & $185$\!\!\!\!\!\!&$\pm$&\!\!\!\!\!\!$ 3$&PDG \cite{PDG} \\
$f_0(1370)$ &  $1475$ \!\!\!\!\!\!\!\!&$\pm$& \!\!\!\!\!\!$ 6$& $113$\!\!\!\!\!\!&$\pm$&\!\!\!\!\!\!$ 11$ &\lhcb \cite{LHCb:2012ae}\\
$f_0(1500)$ & $1505$ \!\!\!\!\!\!\!\!&$\pm$& \!\!\!\!\!\!$ 6$ & $109$\!\!\!\!\!\!&$\pm$&\!\!\!\!\!\!$7$&PDG \cite{PDG}\\
$f_2'(1525)$ & $1525$ \!\!\!\!\!\!\!\!&$\pm$& \!\!\!\!\!\!$ 5$ & $73$\!\!\!\!\!\!&$\pm$&\!\!\!\!\!\!$ 6$&PDG \cite{PDG} \\
$f_2(1640)$ & $1639$ \!\!\!\!\!\!\!\!&$\pm$& \!\!\!\!\!\!$ 6$ & $99$\!\!\!\!\!\!&$\pm$&\!\!\!\!\!\!$ 60$&PDG \cite{PDG} \\
$\phi(1680)$& $1680$ \!\!\!\!\!\!\!\!&$\pm$& \!\!\!\!\!\!$20$ &$150$\!\!\!\!\!\!&$\pm$&\!\!\!\!\!\!$ 50$&PDG \cite{PDG}\\ 
$f_0(1710)$ & $1720$ \!\!\!\!\!\!\!\!&$\pm$& \!\!\!\!\!\!$ 6$ & $135$\!\!\!\!\!\!&$\pm$&\!\!\!\!\!\!$ 8$ &PDG \cite{PDG}\\
$f_2(1750)$ & $1737$ \!\!\!\!\!\!\!\!&$\pm$& \!\!\!\!\!\!$ 9$ & $151$\!\!\!\!\!\!&$\pm$&\!\!\!\!\!\!$ 33$ &\belle \cite{Abe:2003vn}\\
$f_2(1950)$ & $1980$ \!\!\!\!\!\!\!\!&$\pm$& \!\!\!\!\!\!$14$ & $297$\!\!\!\!\!\!&$\pm$&\!\!\!\!\!\!$ 13$&\belle \cite{Abe:2003vn} \\
\hline
\end{tabular}\label{tab:resparam}
\end{table}

The $f_0(980)$ is described by a Flatt\'e resonance shape, see Eq.~(\ref{Eq:Flatte}). The parameters describing the function are the mass, and the couplings $g_{\pi\pi}$ and $g_{KK}$, which are fixed in the fit from the previous analysis of $\Bsb \to J/\psi\pip\pim$ ~\cite{LHCb:2012ae}. The parameters are  $m_0=939.9\pm 6.3$ MeV, $g_{\pi\pi}=199\pm 30$ MeV and $g_{KK}/g_{\pi\pi}=3.0\pm 0.3$. All background and efficiency parameters are fixed in the fit.  


To determine the complex amplitudes in a specific model, the data are fitted maximizing the unbinned likelihood given as
\begin{equation}
\mathcal{L}=\prod_{i=1}^{N}F(s^i_{12},s^i_{23},\theta^i_{J/\psi}),
\end{equation} 
where $N$ is the total number of candidates, and $F$ is the total PDF defined in Eq.~(\ref{eq:pdf}). The PDF normalization is accomplished by first normalizing the $J/\psi$ helicity dependent part by analytical integration, and then for the mass dependent part using numerical integration over 400$\times$800 bins. 

The fit determines the relative values of the amplitude strengths, $a_\lambda^{R_i}$, and phases, $\phi_\lambda^{R_i}$, defined in Eq.~(\ref{amplitude-eq}). We choose to fix $a_0^{\phi(1020)}=1$. 
As only relative phases are physically meaningful, one phase in each helicity grouping must be fixed. In addition, because $J/\psi \KpKm$ is a self-charge-conjugate mode and does not determine the initial $B$ flavor, the signal function is an average of $\Bs$ and $\Bsb$. If we consider no $\KpKm$ partial-waves of a higher order than D-wave, then we can express the differential decay rate ($d\Gamma/dm_{KK}\,d\cos\theta_{KK\,}d\cos\theta_{\jpsi}$) derived from Eq.~(\ref{amplitude-eq}) in terms of S-, P-, and D-waves including helicity 0 and $\pm1$ components. The differential decay rates for $\Bsb$ and $\Bs$, respectively are
\def \A {{\cal A}}
\begin{align}
&\hspace{-8mm}\frac{d\overline\Gamma}{dm_{KK}\,d\cos\theta_{KK\,}d\cos\theta_{\jpsi}}=\nonumber\\
&\hspace{4mm} \left|\A^s_{S_0}e^{i\phi^s_{S_0}}+\A^s_{P_0} e^{i\phi^s_{P_0}}\cos \theta_{KK}+\A^s_{D_0} e^{i\phi^s_{D_0}} \left(\frac{3}{2}\cos^2 \theta_{KK} -\frac{1}{2}\right)\right|^2\sin^2\theta_{J/\psi}\nonumber\\
&+\left|\A^s_{P_{\pm 1}} e^{i\phi^s_{P_{\pm 1}}} \sqrt{\frac{1}{2}}\sin\theta_{KK} +\A^s_{D_{\pm 1}}e^{i\phi^s_{D_{\pm 1}}} \sqrt{\frac{3}{2}}\sin\theta_{KK} \cos\theta_{KK}\right|^2 \frac{1+\cos^2\theta_{J/\psi}}{2},\label{Rb}
\end{align}
and
\begin{align}
&\hspace{-8mm}\frac{d\Gamma}{dm_{KK}\,d\cos\theta_{KK\,}d\cos\theta_{\jpsi}}=\nonumber\\
&\hspace{4mm} \left|\A^s_{S_0}e^{i\phi^s_{S_0}}-\A^s_{P_0} e^{i\phi^s_{P_0}}\cos \theta_{KK}+\A^s_{D_0} e^{i\phi^s_{D_0}}\left(\frac{3}{2}\cos^2 \theta_{KK} -\frac{1}{2}\right)\right|^2\sin^2\theta_{J/\psi}\nonumber\\
&+\left|\A^s_{P_{\pm 1}} e^{i\phi^s_{P_{\pm 1}}} \sqrt{\frac{1}{2}}\sin\theta_{KK} -\A^s_{D_{\pm 1}}e^{i\phi^s_{D_{\pm 1}}} \sqrt{\frac{3}{2}}\sin\theta_{KK} \cos\theta_{KK}\right|^2 \frac{1+\cos^2\theta_{J/\psi}}{2},\label{R}
\end{align}
where $\A^s_{k_\lambda}$ and $\phi^s_{k_\lambda}$ are the sum of amplitudes and reference phases, for the spin-$k$ resonance group, respectively.
The  decay  rate  for $\Bs$ is similar to that of $\Bsb$, except $\theta_{\KpKm}$ and $\theta_{J/\psi}$ are now changed to $\pi-\theta_{\KpKm}$ and $\pi-\theta_{J/\psi}$ respectively, as a result of using $K^-$ and $\mu^-$ to define the helicity angles and hence the signs change in front of the $\A^s_{P_0}$ and $\A^s_{D_{\pm 1}}$ terms.

Summing Eqs.~(\ref{Rb}) and (\ref{R}) results in cancellation of the interference  involving $\lambda=0$ terms for spin-1, and the $\lambda=\pm1$ terms for spin-2, as they appear with opposite signs for  $\Bsb$ and $\Bs$ decays.  Therefore we have to fix one phase in the spin-1 ($\lambda=0$) group ($\phi^s_{P_0}$) and one in the spin-2 ($\lambda=\pm1$) group ($\phi^s_{D_{\pm 1}}$). The other phases in each corresponding group are determined relative to that of the fixed resonance.


\subsection{Fit results}
\label{sec:fitresult}
The goodness of fit is calculated from 3D partitions of $s_{12}$, $s_{23}$ and $\cos\theta_{\jpsi}$. We use the Poisson likelihood $\chi^2$ \cite{Baker:1983tu} defined as 
\begin{equation}
\chi^2=2\sum_{i=1}^N\left[  x_i-n_i+n_i \text{ln}\left(\frac{n_i}{x_i}\right)\right],
\end{equation}
where $n_i$ is the number of candidates in the three dimensional bin $i$ and $x_i$ is the expected number of candidates in that bin according to the fitted likelihood function. An adaptive binning algorithm is used, requiring a minimum of 25 entries in each bin. 
The associated number of degrees of freedom (ndf) is $N-k-1$, where $k$ is the number of free parameters in the likelihood function. The $\chi^2/\text{ndf}$ and the negative of the logarithm of the likelihood, $\rm -ln\mathcal{L}$, of the fits are given in Table~\ref{tab:RMchi2}. Starting values of parameters are varied in order to ensure that global likelihood minimums are found rather than local minimums.
\begin{table}[h!t!p!]
\centering
\caption{$\chi^2/\text{ndf}$ and $\rm -ln\mathcal{L}$ of different resonance models.}
\vspace{0.2cm}
\begin{tabular}{cccc}
\hline
Resonance model & $\rm -ln\mathcal{L}$& $\chi^2/\text{ndf}$ \\
\hline
Best fit  & 29,275 & 649/545~=~1.1908\\
Best fit + $f_2(1270)$ & 29,273  &  644/541~=~1.1911\\
Best fit + $f_0(1500)$ & 29,274  &  647/543~=~1.1915\\
\hline
\end{tabular}
\label{tab:RMchi2}
\end{table}

\begin{figure}[hbt]
\centering
\includegraphics[scale=0.65]{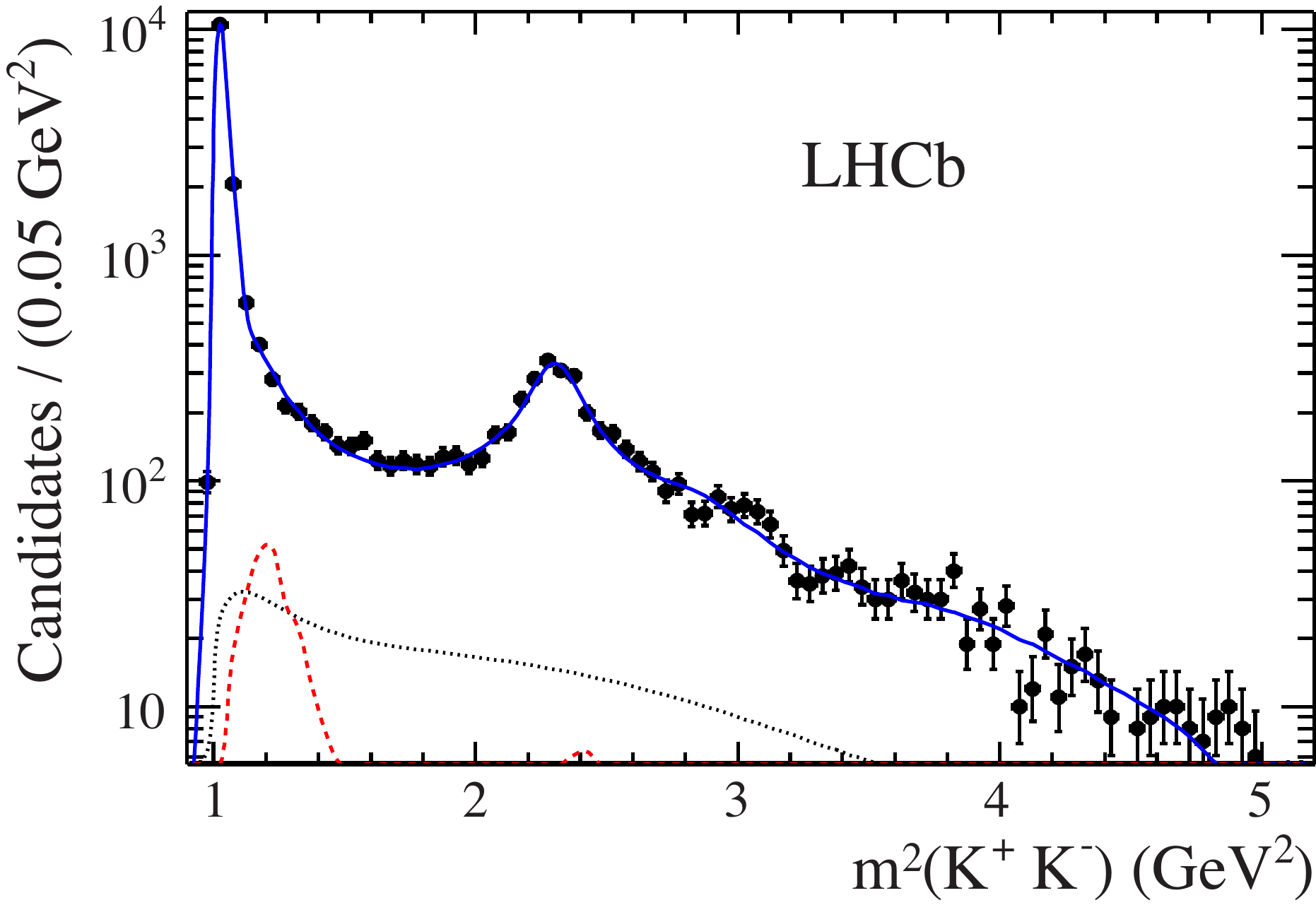}
\vspace{-2mm}
\caption{\small Dalitz plot fit projection of  $ m^2(K^+ K^-)$ using a logarithmic scale. The points with error bars are data, the (black) dotted curve shows the combinatorial background, the (red) dashed curve indicates the reflection from the misidentified $\Bdb\to J/\psi\Km\pip$ decays, and the (blue) solid line represents the total.}
\label{fig:RMfit1}
\end{figure}

\begin{figure}[b]
\centering
\vspace{2.5mm}
\includegraphics[scale=0.4]{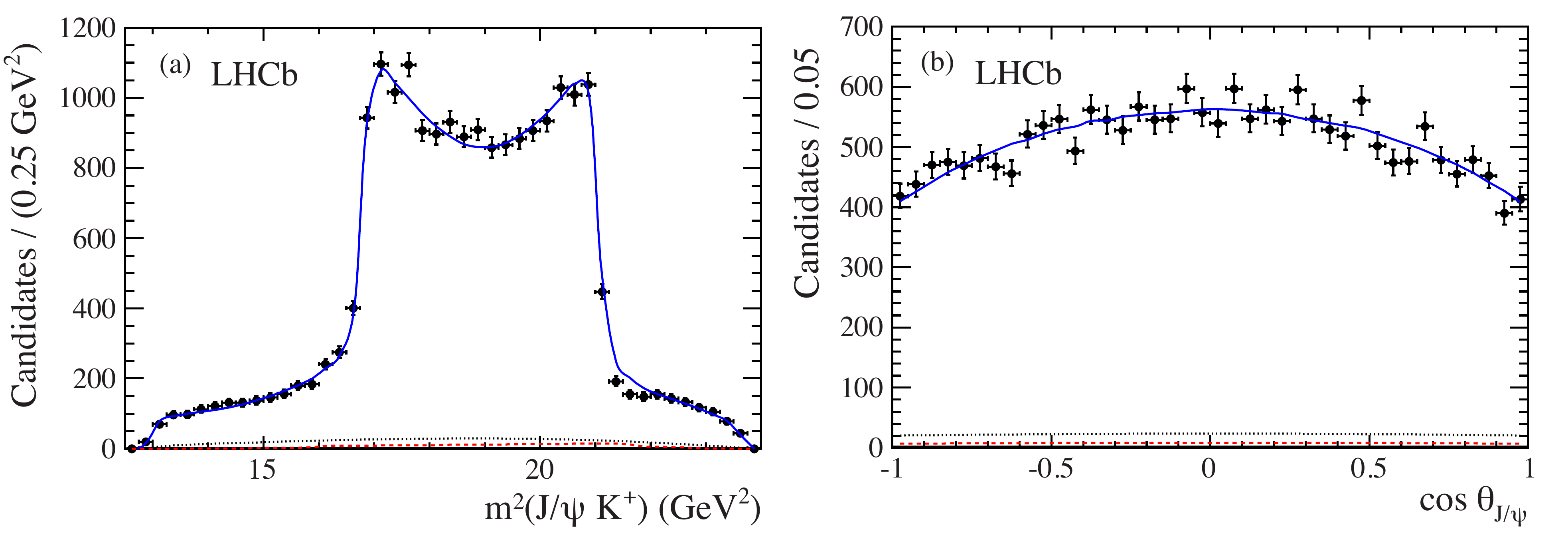}
\vspace*{-2mm}
\caption{\small Dalitz plot fit projections of (a) $ m^2(J/\psi K^{+})$ and (b) $\cos \theta_{J/\psi}$. The points with error bars are data, the (black) dotted curve shows the combinatorial background, the (red) dashed curve indicates the reflection from the misidentified $\Bdb\to J/\psi\Km\pip$ decays, and the (blue) solid line represents the total fit results.}
\label{fig:RMfit2}
\vspace*{10mm}
\end{figure}

\begin{figure}[htb]
\centering
\includegraphics[scale=0.72]{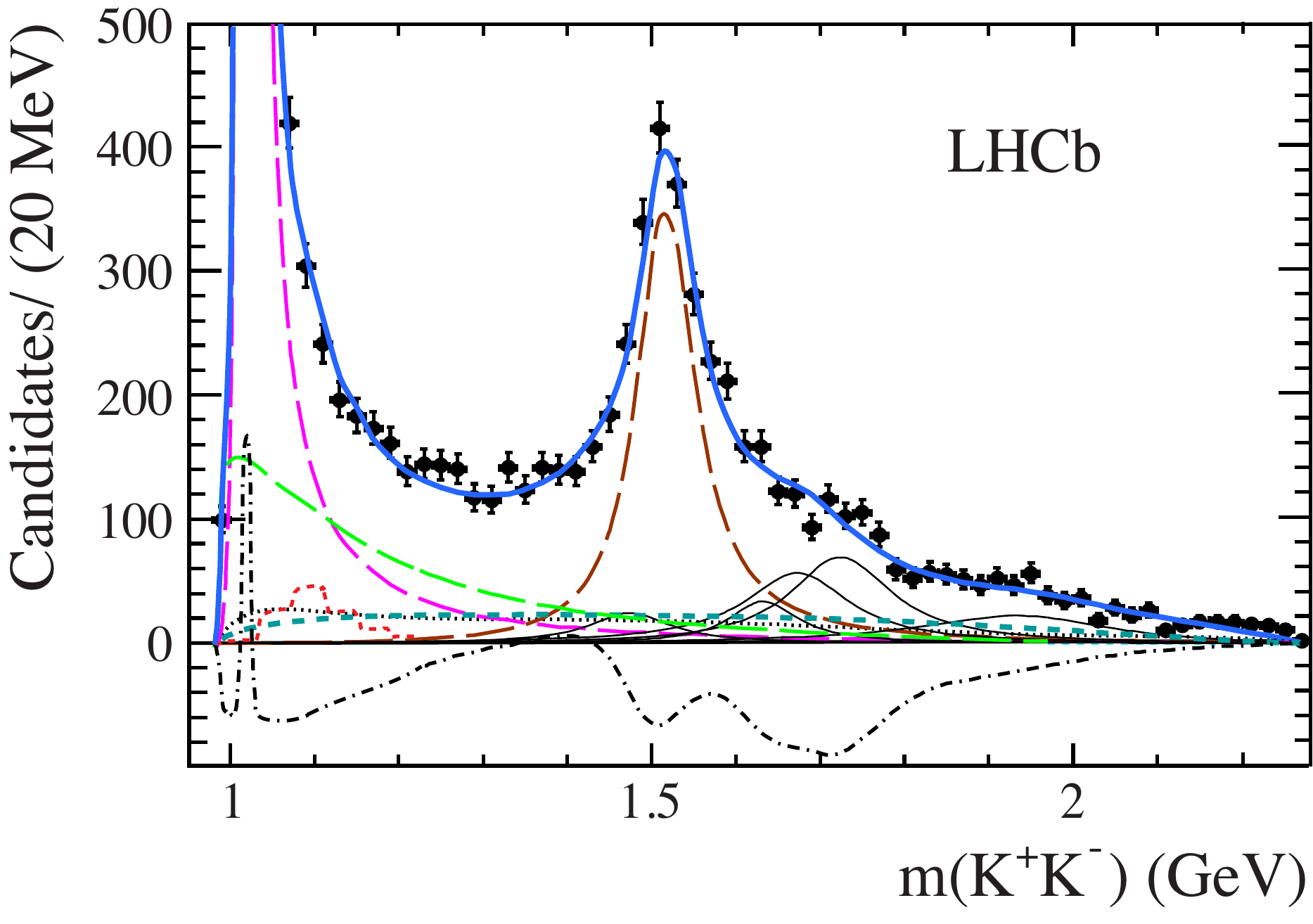}
\vspace*{-3mm}
\caption{\small Dalitz fit projection of  $m(K^+ K^-)$. The points represent the data, the dotted (black) curve shows the combinatorial background, and the dashed (red) curve indicates the reflection from misidentified $\Bdb\to J/\psi\Km\pip$ decays. The largest three resonances $\phi(1020)$, $f_2'(1525)$ and $f_0(980)$ are shown by magenta, brown and green long-dashed curves, respectively; all other resonances are shown by thin black curves.  The dashed (cyan) curve is the non-resonant contribution. The dot-dashed (black) curve is the contribution from the interferences, and the solid (blue) curve represents the total fit result.}
\label{fig:RMfit3} 
\end{figure}

Attempts to add one more resonance  such as $f_2(1270)$ and $f_0(1500)$  improve the $-\rm ln\mathcal{L}$ marginally, but  the $\chi^2/\rm ndf$ are worse than the best fit model. 
We retain only those resonances that are more than $2\sigma$ significant,
except for the $f_2(1750)$  where we allow the $|\lambda| = 1$ component,  since the $\lambda = 0$ component is significant. 
For models with one more resonance, the additional components never have more than 2$\sigma$  significance. 
Figure~\ref{fig:RMfit1} shows the projection of $m^2(K^+K^-)$ for the best fit model, the $m^2(J/\psi K^+)$ and $\cos \theta_{J/\psi}$ projections are displayed in Fig.~\ref{fig:RMfit2}. The projection of the $\Kp\Km$ invariant mass spectrum is shown in Fig.~\ref{fig:RMfit3}.

While a complete description of the $\Bsb\to\jpsi K^+K^-$ decay is given in terms of the fitted amplitudes and phases, knowledge of the contribution of each component can be
summarized by defining a fit fraction, ${\cal{F}}^R_{\lambda}$. To determine ${\cal{F}}^R_{\lambda}$ we integrate the squared amplitude of $R$ over the Dalitz plot. The yield is then normalized by integrating the entire signal function over the same area. Specifically,  
\begin{equation}
{\cal{F}}^R_{\lambda}=\frac{\int\left| a^R_{\lambda} e^{i\phi^R_{\lambda}} \mathcal{A}_{\lambda}^{R}(s_{12},s_{23},\theta_{J/\psi})\right|^2 ds_{12}\;ds_{23}\;d\cos\theta_{J/\psi}}{\int S(s_{12},s_{23},\theta_{J/\psi})  ~ds_{12}\;ds_{23}\;d\cos\theta_{J/\psi}}.
\end{equation}
Note that the sum of the fit fractions is not necessarily unity due to the potential presence of interference between two resonances. Interference term fractions are given by
\begin{equation}
\label{eq:inter}
{\cal{F}}^{RR'}_{\lambda}=\mathcal{R}e\left(\frac{\int a^R_{\lambda}\; a^{R'}_{\lambda}e^{i(\phi^R_{\lambda}-\phi_{\lambda}^{R'})} \mathcal{A}_{\lambda}^{R}(s_{12},s_{23},\theta_{J/\psi}) {\mathcal{A}_{\lambda}^{R'}}^{*}(s_{12},s_{23},\theta_{J/\psi}) ds_{12}\;ds_{23}\;d\cos\theta_{J/\psi}}{\int S (s_{12},s_{23},\theta_{J/\psi}) ~ds_{12}\;ds_{23}\;d\cos\theta_{J/\psi}}\right),
\end{equation}
and
\begin{equation}
\sum_{\lambda}\left(\sum_R {\cal{F}}^R_{\lambda}+\sum^{R\neq R'}_{RR'} {\cal{F}}^{RR'}_{\lambda}\right) =1.
\end{equation}
If the Dalitz plot has more destructive interference than constructive interference, the sum of the fit fractions will be greater than unity. Conversely, the sum will be less than one if the Dalitz plot exhibits constructive interference. Note that interference  between different spin-$J$ states vanishes because the $d^J_{\lambda0}$ angular functions in $\mathcal{A}^R_{\lambda}$ are orthogonal.
 
The determination of the statistical uncertainties of the fit fractions is difficult because they depend on the statistical uncertainty of every fitted magnitude and phase. Therefore we determine the uncertainties from simulated experiments. We perform 500 experiments: each sample is generated according to the model PDF, input parameters are taken from the fit to the data.  The correlations of fitted parameters are also taken into account. For each  experiment the fit fractions are calculated. 
The distributions of the obtained fit fractions are described by Gaussian functions. The r.m.s. widths of the Gaussian functions are taken as the statistical uncertainties on the corresponding parameters. The fit fractions and phases of the contributing components are given in Table~\ref{tab:fitfraction}, while the  fit fractions of the interference terms are quoted in Table~\ref{tab:interferencefraction}. 

\renewcommand{\arraystretch}{1.2}
\begin{table}[hbt]
\centering
\caption{Fit fractions (\%) and phases of contributing components. For P- and D-waves $\lambda$ represents the  helicity.}
\vspace{0.2cm}
\begin{tabular}{lcc}
\hline
Component & Fit fraction (\%) & Phase (degree)\\
\hline
$\phi(1020)$, $\lambda=0$   & $32.1\pm 0.5 \pm 0.8$ & 0(fixed) \\
$\phi(1020)$, $|\lambda|=1$ & $34.6\pm 0.5 \pm 1.3$ & 0(fixed) \\
$f_0(980)$                  & $12.0\pm 1.8^{+2.8}_{-2.5}$  & $-294\pm 8 \pm 25$\\
$f_0(1370)$                 & $1.2\pm 0.3^{+0.3}_{-1.2}$  & $-81\pm 8\pm 8$\\
$f_2'(1525)$, $\lambda=0$   & $9.9\pm 0.7^{+2.4}_{-1.6}$  & 0(fixed)\\
$f_2'(1525)$, $|\lambda|=1$ & $5.1\pm 0.9^{+1.8}_{-1.4}$  & 0(fixed)\\
$f_2(1640)$, $|\lambda|=1$  & $1.5\pm 0.7^{+0.7}_{-0.9}$  & $165\pm 27^{+13}_{-44}$\\
$\phi(1680)$, $|\lambda|=1$ & $3.4\pm 0.3^{+4.4}_{-0.3}$  & $106\pm 14^{+260}_{-210}$\\
$f_2(1750)$, $\lambda=0$    & $2.6 \pm 0.5^{+1.0}_{-0.6}$ & $238\pm 8 \pm
9$\\
$f_2(1750)$, $|\lambda|=1$  & $1.8\pm 1.0^{+2.2}_{-1.8}$  & $45\pm 30^{+16}_{-70}$\\        
$f_2(1950)$, $\lambda=0$    & $0.4 \pm 0.2^{+0.2}_{-0.4}$ & $46\pm 17^{+110}_{-~20}$\\
$f_2(1950)$, $|\lambda|=1$  & $1.7\pm 0.5^{+2.5}_{-1.7}$  & $-53\pm 26^{+150}_{-~80}$\\
Non-resonant                 & $6.0\pm 1.6^{+2.0}_{-2.2}$  & $-39\pm 6\pm 23$\\
\hline
Total                         &$112.3 $\\
\hline
\end{tabular}
\label{tab:fitfraction}
\end{table}
\renewcommand{\arraystretch}{1.0}


\begin{sidewaystable}[h!t!p!]
\centering
\caption{\small  Fit fractions matrix for the best fit in units of \%. The diagonal elements correspond to the decay fractions in Table~\ref{tab:fitfraction}. The off-diagonal elements give the fit fractions of the interference. The null values originate from the fact that any interference contribution between different spin-$J$  state integrates to zero. Here the resonances are labeled by their masses in MeV and the subscripts denote the helicities.}
\vspace{0.2cm}
\begin{tabular}{clccccccccccccc}
\hline
&$1020_0$&$1020_1$&980&1370&$1525_0$& $1525_1$&$1640_1$&$1680_1$&$1750_0$&$1750_1$& $1950_0$&$1950_1$& NR\\
\hline
$1020_0$ &32.1& 0 &0&0&0&0&0&0&0&0&0&0&0\\
$1020_1$ &&34.6&0&0&0&0&0&2.0&0&0&0&0&0\\
~980& & &12.0&-2.3&0&0&0&0&0&0&0&0&-4.7\\ 
1370&&&&1.2&0&0&0&0&0&0&0&0&1.5\\   
$1525_0$&&&&&9.9&0&0&0&-4.5&0&0.9&0&0\\
$1525_1$&&&&&&5.1&-0.9&0&0&2.5&0&-1.8&0\\
$1640_1$&&&&&&&1.5&0&0&-2.4&0&0.7&0\\
$1680_1$&&&&&&&&3.4&0&0&0&0&0\\
$1750_0$&&&&&&&&&2.6&0&-0.4&0&0\\
$1750_1$&&&&&&&&&&1.8&0&-2.2&0\\
$1950_0$&&&&&&&&&&&0.4&0&0\\
$1950_1$&&&&&&&&&&&&1.7&0\\
NR      &&&&&&&&&&&&&6.0\\
\hline
\end{tabular}
\label{tab:interferencefraction}
\end{sidewaystable}

Table~\ref{tab:comparison} shows a comparison of the fit fractions when a different parametrization is used for the $f_0(980)$ resonance shape.  The BES $f_0(980)$ functional form is the same as  ours with the parameters $m_0=965\pm 10$ MeV, $g_{\pi\pi}=165\pm 18$ MeV and $g_{KK}/g_{\pi\pi}=4.21\pm 0.33$ \cite{Ablikim:2004wn}. The \babar collaboration assumes that the non-resonant S-wave is small and consistent with zero \cite{Papagallo}.  The \babar functional form is different and parametrized as
\begin{equation}
A_R(s_{23})=\frac{1}{m^2_{R}-s_{23}-im_{R}\Gamma_{R}\rho_{KK}},
\end{equation}
with $\rho_{KK}=2P_R/\sqrt{s_{23}}$. The parameters are $m_R=922\pm 3$ MeV and $\Gamma_{R}=240 \pm 80$ MeV, taken from \babar's Dalitz plot analysis of $D_s^+ \to K^+ K^- \pi^+$~\cite{delAmoSanchez:2010yp}. The $f_0(980)$ fraction is smaller in the \babar parametrization, while the total S-wave fraction is consistent in the three different parameterizations. In all cases, the dominant component is  the  $\phi(1020)$, the second largest contribution is the $f_2'(1525)$, and the third the $f_0(980)$ resonance.  There are also significant contributions from the $f_0(1370)$, $f_2(1640)$, $\phi(1680)$, $f_2(1750)$, $f_2(1950)$ resonances, and non-resonant  final states. The amount of $f_0(980)$ is strongly parametrization dependent, so we treat these three models separately and 
do not assign any systematic uncertainty based on the use of these different $f_0(980)$ shapes.  
Therefore we refrain from quoting a branching fraction measurement for the decay $\Bsb \to J/\psi f_0(980)$.

The determination of the  parameters of the $f_2'(1525)$ resonance are not dependent on the $f_0(980)$ parametrization.  
The parameters of the $f_2'(1525)$ are determined to be:
\begin{eqnarray*}
 m_{f_2'(1525)} &=& 1522.2 \pm 2.8^{+5.3}_{-2.0}~\rm MeV, \\
\Gamma_{f_2'(1525)} &=& 84 \pm 6 ^{+10}_{-~5}~ \rm MeV.
\end{eqnarray*}
Whenever two  or more uncertainties are quoted, the first is the statistical and the second systematic. The latter will be discussed in Section~\ref{sec:sys}. 
These values are the most accurate determinations of the $f_2'(1525)$ resonant parameters~\cite{PDG}. Note that our determination of the mass has the same uncertainty as the current PDG average.

\begin{table}[t]
\centering
\caption{Comparison of the fit fractions (\%) with the LHCb, BES and BaBar $f_0(980)$ parameterizations described in the text. For P- and D-waves, $\lambda$ represents the  helicity.}
\vspace{0.2cm}
\begin{tabular}{lccc}
\hline
Component & \lhcb & BES & \babar\\
\hline
$\phi(1020)$, $\lambda=0$   & $32.1\pm 0.5~ $ & $32.1\pm$0.5~~& $32.0\pm0.5$~  \\
$\phi(1020)$, $|\lambda|=1$ & $34.6\pm 0.5~ $ & $34.6\pm0.5~$&$34.5\pm0.5$~ \\
$f_0(980)$                  & $12.0\pm 1.8~$  & $9.2\pm1.4$&$4.8\pm1.0$\\
$f_0(1370)$                 & $1.2\pm 0.3$  &$1.2\pm0.3 $&$1.3\pm0.3$ \\
$f_2'(1525)$, $\lambda=0$   & $9.9\pm 0.7$  &$9.8\pm0.7$&$9.5\pm0.7$\\
$f_2'(1525)$, $|\lambda|=1$ & $5.1\pm 0.9$  &$5.1\pm0.9$&$4.9\pm0.9$\\
$f_2(1640)$, $|\lambda|=1$  & $1.5\pm 0.7$  &$1.5\pm0.7$&$1.5\pm0.7$\\
$\phi(1680)$, $|\lambda|=1$ & $3.4\pm 0.3$  &$3.4\pm0.3$&$3.4\pm0.3$ \\
$f_2(1750)$, $\lambda=0$    & $2.6 \pm 0.5$ &$2.5\pm0.5$&$2.2\pm0.5$\\
$f_2(1750)$, $|\lambda|=1$  & $1.8\pm 1.0$ &$1.8\pm1.0$&$1.9\pm1.0$\\        
$f_2(1950)$, $\lambda=0$    & $0.4 \pm 0.2$ &$0.4\pm0.2$&$0.4\pm0.2$\\
$f_2(1950)$, $|\lambda|=1$  & $1.7\pm 0.5$&$1.8\pm0.5$&$1.8\pm0.5$\\
Non-resonant S-wave                & $6.0\pm 1.4$&$4.7\pm1.2$&$8.6\pm1.7$\\
\hline
Interference between S-waves &$-$5.5&$-$1.7&$-$1.1\\
Total S-wave                  &13.7&13.4&13.6 \\
\hline
-ln$\mathcal{L}$ & 29,275 & 29,275 & 29,281\\
$\chi^2/\rm ndf$& 649/545&653/545&646/545\\
\hline
\end{tabular}
\label{tab:comparison}
\end{table}
\newpage
\subsection{\boldmath $K^+K^-$ S-wave in the $\phi(1020)$ mass region}
\label{sec:kks-wave}
\begin{figure}[b]
\centering
\includegraphics[scale=0.53]{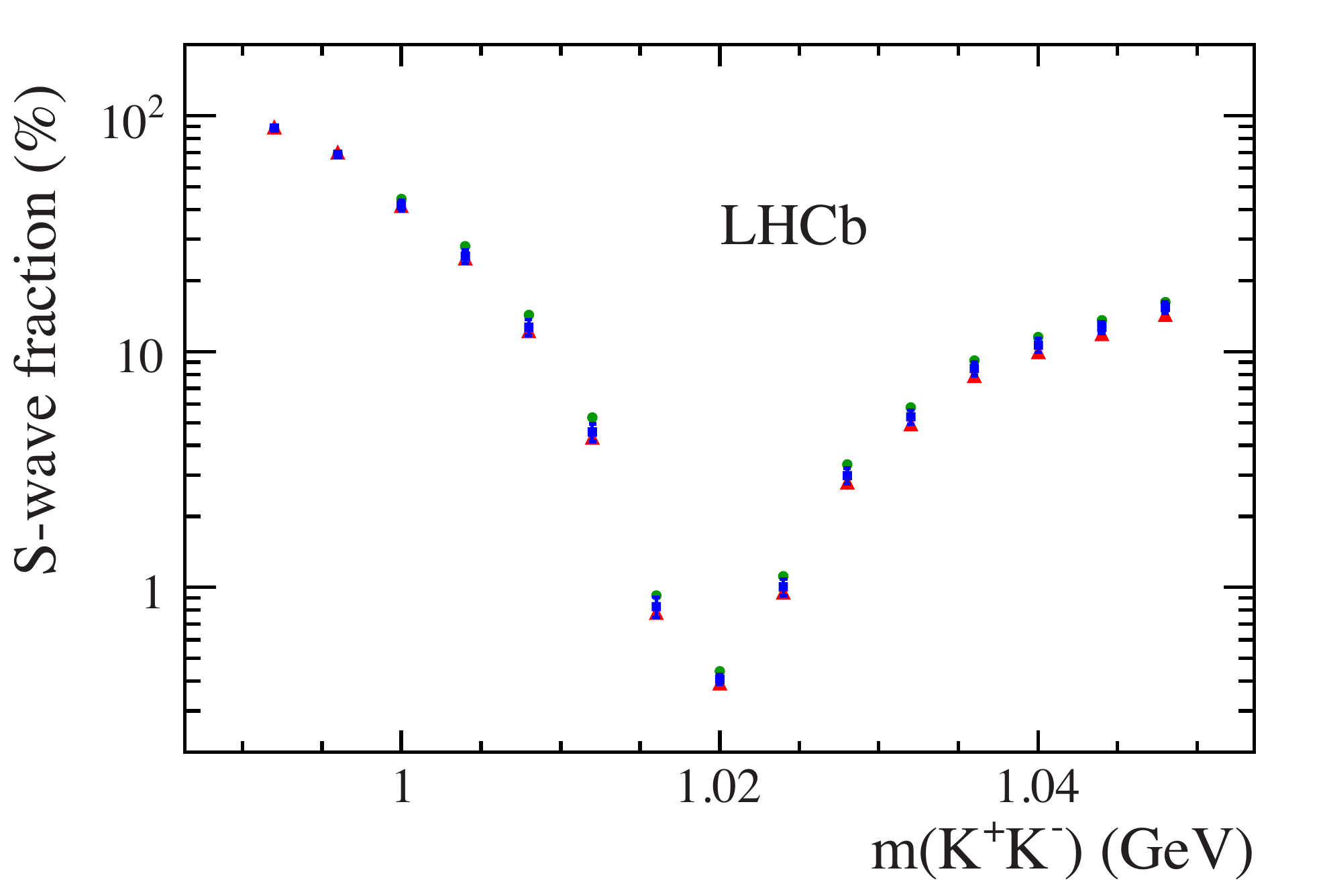}
\vspace*{-2mm}
\caption{\small S-wave fraction as a function of $m(\Kp\Km)$ starting from 990 MeV up to 1050 MeV in 4 MeV mass intervals. The squares (blue), triangles (red), and circles (green) represent the \lhcb, BES and \babar parameterizations of $f_0(980)$, respectively. The experimental statistical uncertainties are only shown for the \lhcb model; they are almost identical for the other cases. The experimental mass resolution is not unfolded.}
\label{fig:Swave}
\end{figure}

\begin{figure}[htb]
\centering
\includegraphics[scale=0.53]{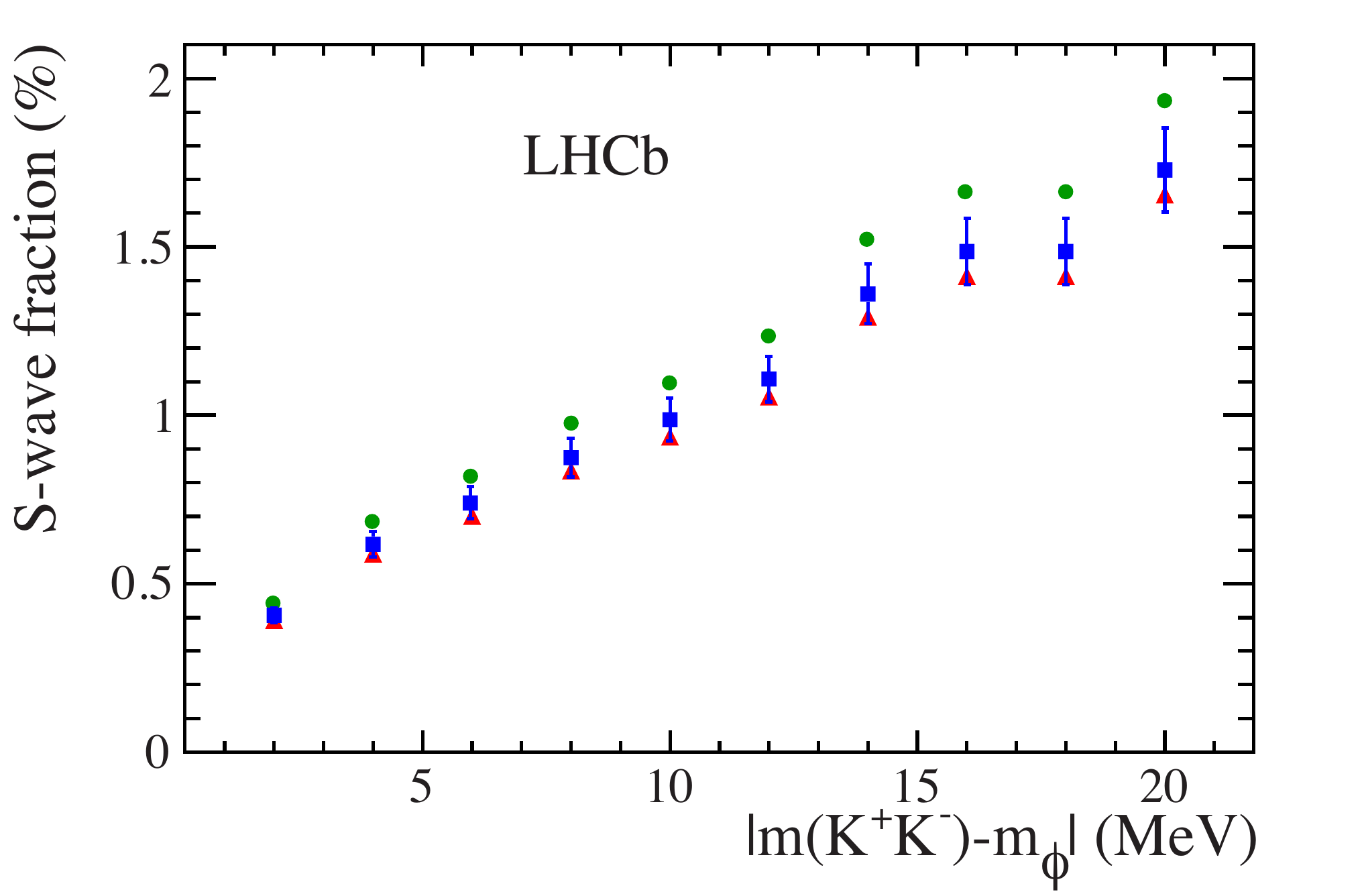}
\caption{\small S-wave fractions  in different $m(\Kp\Km)$ intervals centered on the $\phi$ meson mass. The squares (blue), triangles (red), and circles (green) represent the \lhcb, BES and \babar parameterizations of $f_0(980)$, respectively. The experimental statistical uncertainties are only shown for the \lhcb model; they are almost identical for the other cases. The experimental mass resolution is not unfolded.}
\label{fig:interval}
\end{figure}

It was claimed by Stone and Zhang~\cite{Stone:2008ak} that in the decay of $\Bsb\to J/\psi \phi $, the $K^+K^-$ system can have S-wave contributions under the $\phi(1020)$ peak of order 7\% of the total yield. In order to investigate this possibility we calculate the S-wave fractions as given by the fit in 4~MeV mass intervals between  $990 < m(\KpKm) < 1050$ MeV.  
 The resulting behavior is shown in Fig.~\ref{fig:Swave}.  Here we show the result from our preferred model and also from the alternative $f_0(980)$ parameterizations discussed above. The observation of significant S-wave fractions in this region means that this contribution must be taken into account when measuring \CP violation in the $\phi$ mass region. The total S-wave fraction as a function of the mass interval around the $\phi$ mass is also shown in Fig.~\ref{fig:interval}.  
Using a time dependent analysis of $\Bsb\to J/\psi \phi(1020)$, \lhcb reported  $(2.2\pm 1.2 \pm 0.07)\%$~\cite{LHCb-CONF-2012-002} of  S-wave  within  $\pm 12$ MeV of the $\phi(1020)$ mass peak.  We measure the S-wave fraction within the same mass window as a consistent, and more precise $(1.1\pm 0.1 ^{+0.2}_{-0.1})\%$.  CDF measured the S-wave fraction as $(0.8\pm 0.2)$\% for $m(K^+K^-)$ within about $\pm$9.5 MeV of the $\phi$ mass \cite{Aaltonen:2012ie}, while ATLAS quotes (2$\pm$2)\% for an 11~MeV interval \cite{:2012fu}. These results are consistent with ours. The D0 collaboration, however, claimed a (14.7$\pm$3.5)\% S-wave fraction within approximately $\pm$10 MeV of the $\phi$ meson mass \cite{Abazov:2011ry}, in disagreement with all of the other results.


\subsection{Helicity angle distributions}
\label{sec:helicity}
\begin{figure}[hbt]
\centering
\vspace*{6mm}\hspace*{-4mm} \includegraphics[scale=0.4]{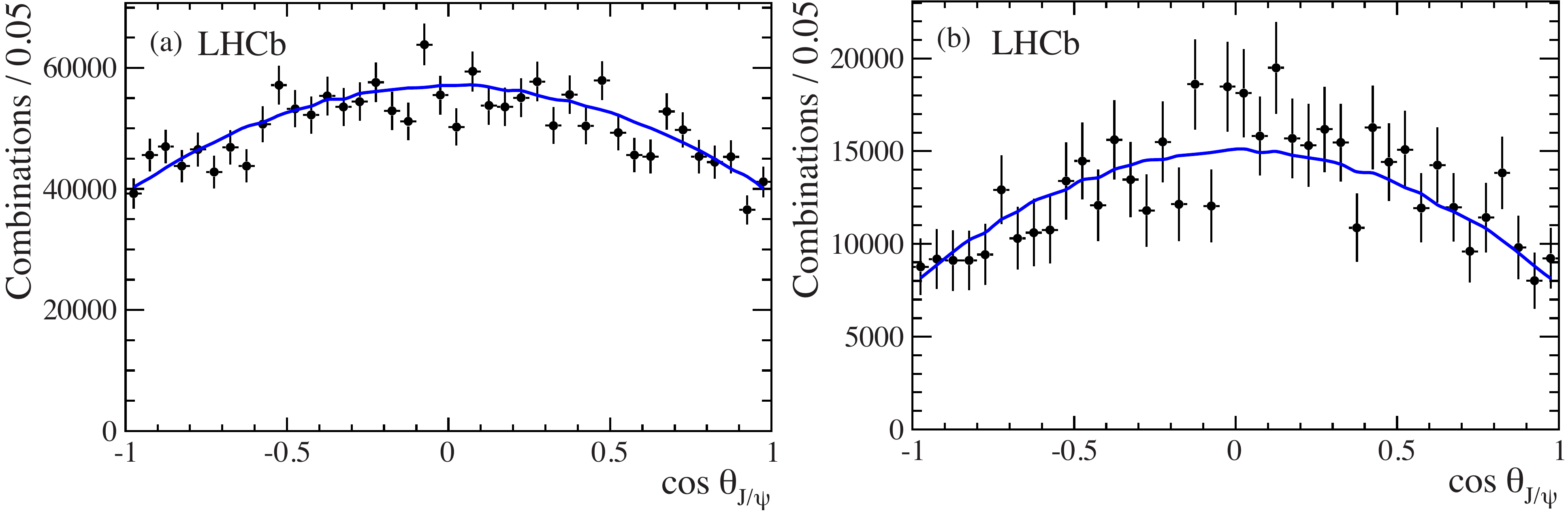}
\caption{\small Background-subtracted efficiency-corrected $\cos \theta_{J/\psi}$ helicity distributions: (a) in $\phi(1020)$ mass region ($\chi^2/\rm ndf =54.4/40$), (b) in $f_2'(1525)$ mass region ($\chi^2/\rm ndf =34.4/40$).}
\label{fig:bgsubhelicosH}
\end{figure}

\begin{figure}[hbt]
\centering
\includegraphics[scale=0.395]{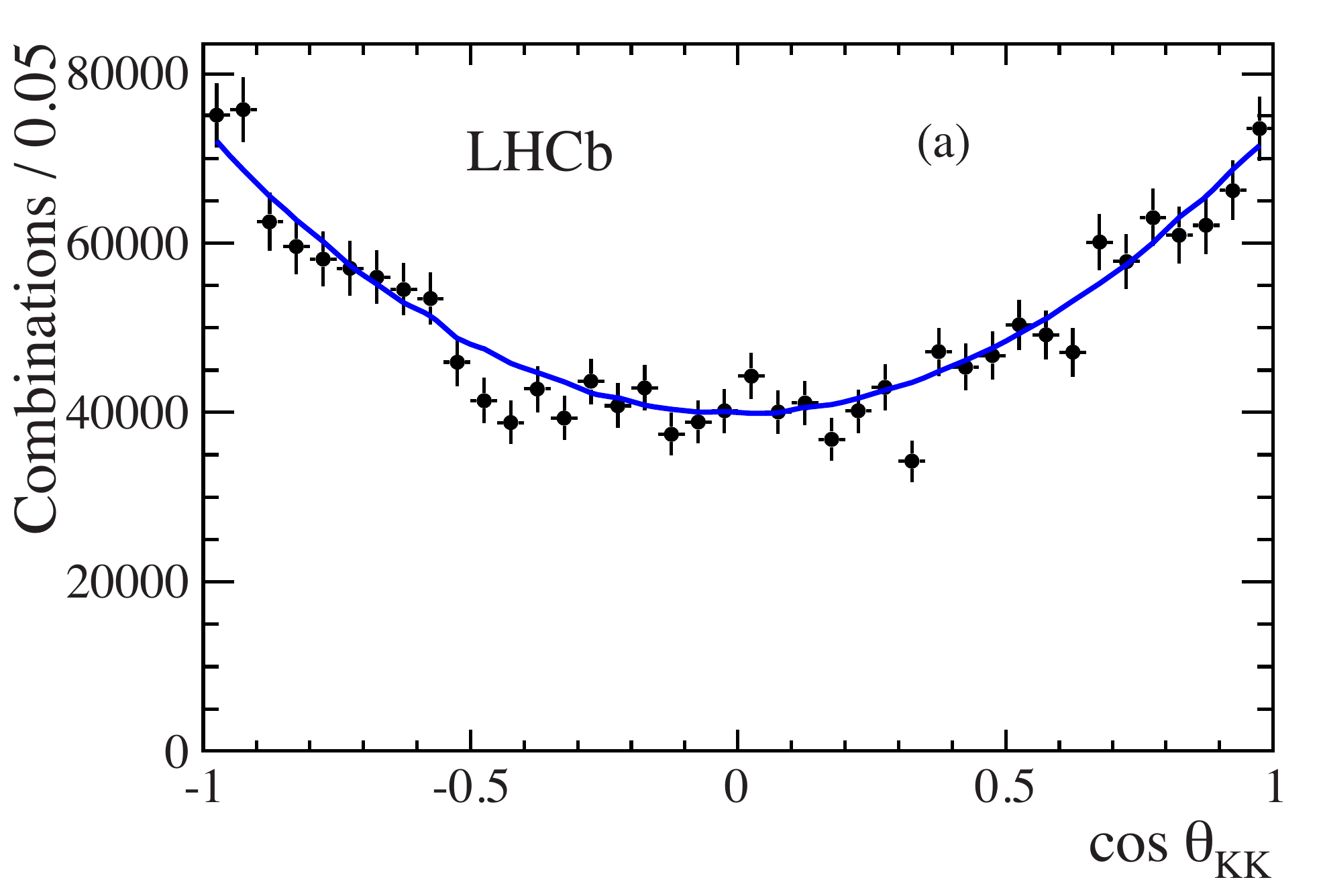}
\includegraphics[scale=0.395]{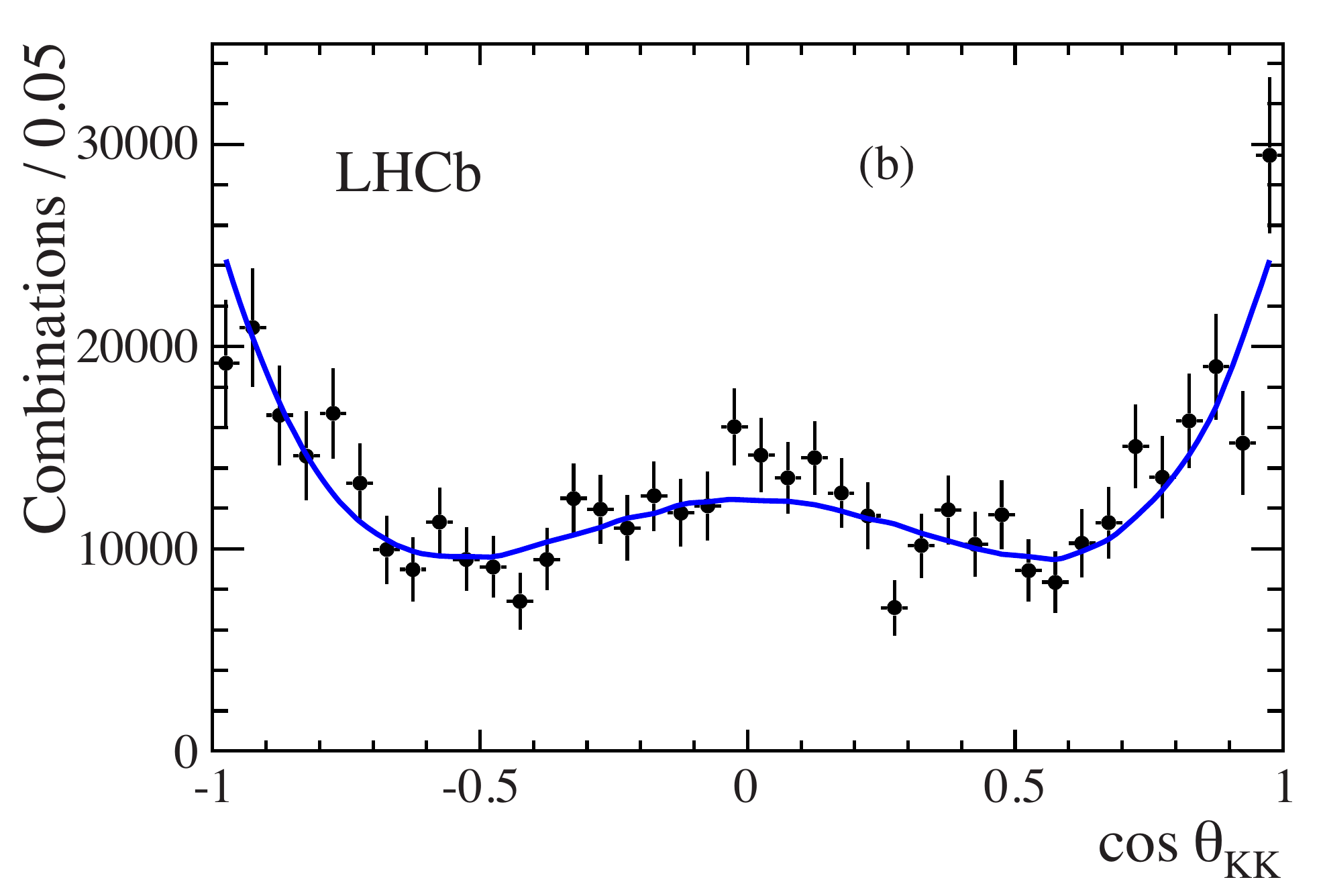}
\caption{\small Background-subtracted efficiency-corrected $\cos \theta_{KK}$ helicity distributions: (a) in $\phi(1020)$ mass region ($\chi^2/\rm ndf =57.4/40$), (b) in $f_2'(1525)$ mass region ($\chi^2/\rm ndf =43.4/40$). The distributions are
compatible with expectations for spin-1 and spin-2, respectively.}
\label{fig:bgsubhelicosK}
\end{figure}

The decay angular distributions or the helicity angle distributions are already included in the signal model via Eqs.~(\ref{heli1}) and (\ref{heli2}). In order to test the fit model we examine the $\cos \theta_{J/\psi}$ and $\cos \theta_{KK}$ distributions in two different $K^+K^-$ mass regions: one is the $\phi(1020)$ region defined within $\pm 12$ MeV of the $\phi(1020)$ mass peak and the other is defined within one full width of the $f_2'(1525)$ mass. 
The background-subtracted efficiency-corrected distributions are shown in Figs.~\ref{fig:bgsubhelicosH} and \ref{fig:bgsubhelicosK}. The distributions are in good agreement with the fit model. 

\subsection{Angular moments}
\label{sec:moments}
The angular moment distributions provide an additional way of
visualizing the effects of different resonances and their interferences, similar to a partial wave analysis. This technique has been used in previous studies~\cite{LHCb:2012ae, Lees:2012kxa}.

\def \t {\theta_{KK}}
\def \S {{\cal S}}
\def \P {{\cal P}}
\def \D {{\cal D}}
We define the angular moments $\langle Y_l^0\rangle$ as the efficiency-corrected and background-subtracted $\Kp\Km$ invariant mass distributions, weighted by orthogonal and normalized spherical harmonic functions $Y_l^0(\cos\t)$,
\begin{equation}
\langle Y_l^0 \rangle = \int_{-1}^{1}d\Gamma(m_{KK},\cos\t)Y_l^0(\cos\t)d\cos\t.
\end{equation}

\begin{figure}[t]
\centering
\hspace{-5mm}\includegraphics[scale=1.05]{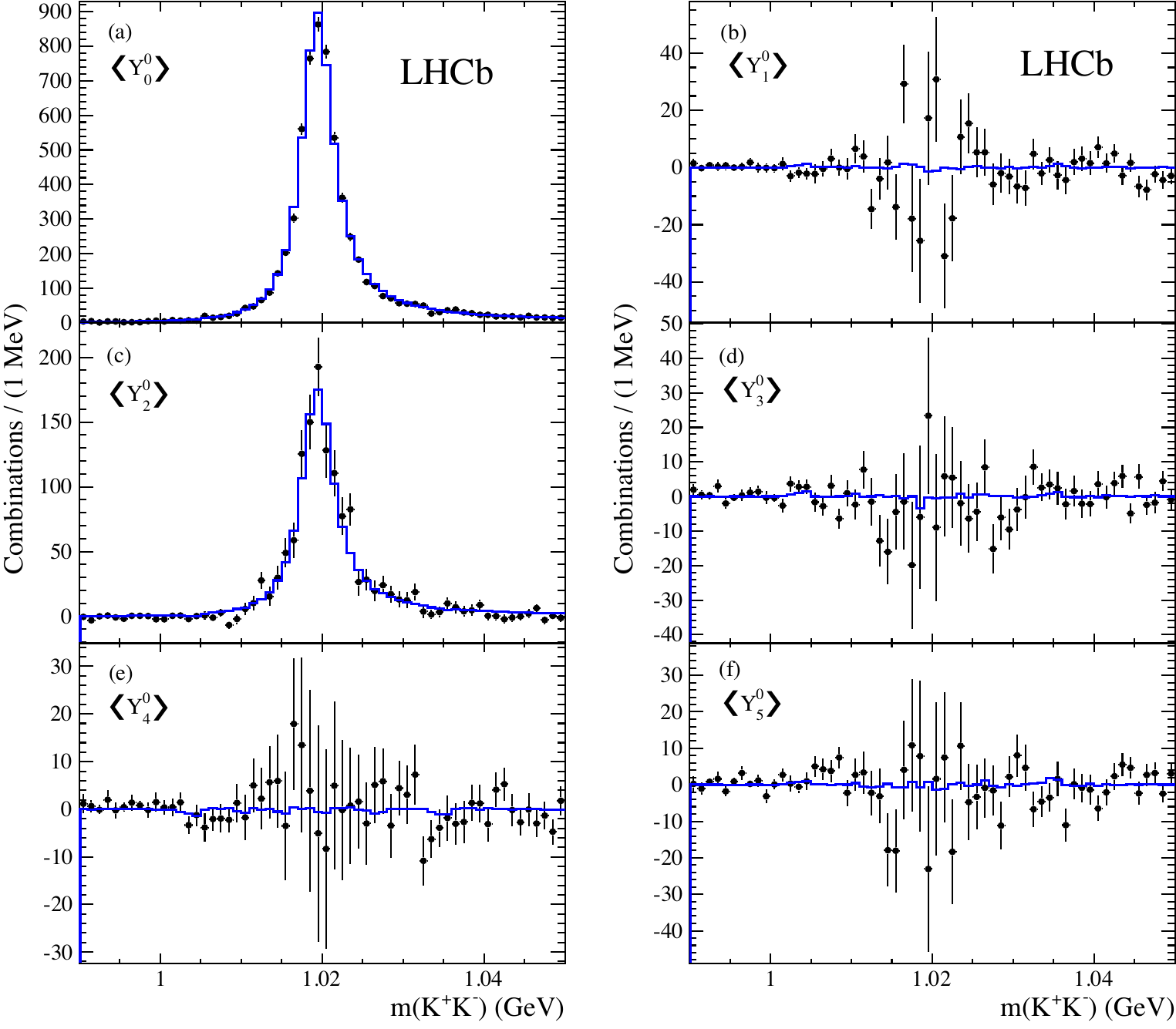}
\caption{\small Dependence of the spherical harmonic moments of $\cos \theta_{KK}$  as a function of the $K^+K^-$ mass  around the $\phi(1020)$ mass peak after efficiency corrections and background subtraction.
The points with error bars are the data and the solid curves are derived from the fit model.}
\label{fig:Sph_phi}
\end{figure}

\begin{figure}[t]
\centering
\vspace{2mm}
\hspace{-5mm}\includegraphics[scale=1.05]{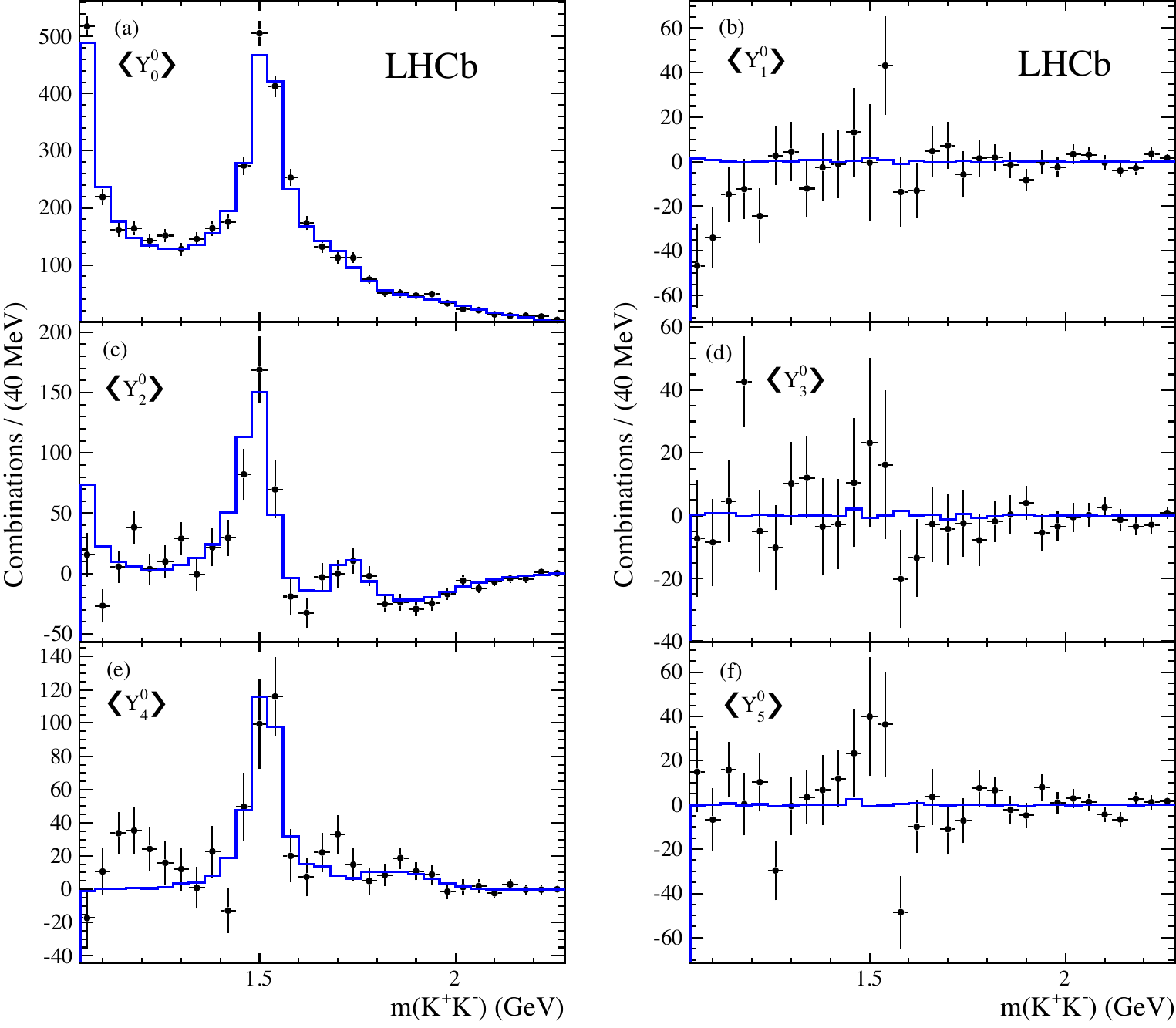}
\caption{\small Dependence of the spherical harmonic moments of $\cos \theta_{KK}$  as a function of the $K^+K^-$ mass above 1050~MeV, after efficiency corrections and background subtraction.
The points with error bars are the data and the solid curves are derived from the fit model.}
\label{fig:Sph}
\end{figure}

If we assume that no $\Kp\Km$ partial-waves of a higher order than D-wave contribute, then we can express the differential decay rate, derived from Eq.~(\ref{amplitude-eq}) in terms of S-, P-, and D-waves including helicity 0 and $\pm1$ components as
\begin{align}
&\hspace{-14mm}\frac{d\overline\Gamma}{dm_{KK}\,d\cos\t}=\nonumber \\
&\hspace{4mm} 2\pi\left|\S_0 Y_0^0(\cos \t)+\P_0 e^{i\Phi_{P_0}}Y_1^0(\cos \t)+\D_0 e^{i\Phi_{D_0}} Y_2^0(\cos \t)\right|^2\nonumber\\
&+2\pi\left|\P_{\pm 1} e^{i\Phi_{P_{\pm 1}}} \sqrt{\frac{3}{8\pi}}\sin\t +\D_{\pm 1}e^{i\Phi_{D_{\pm 1}}} \sqrt{\frac{15}{8\pi}}\sin\t \cos\t\right|^2,
\end{align}
where $\S_{\lambda}$, $\P_{\lambda}$, $\D_{\lambda}$ and $\Phi_{k_\lambda}$ are real-valued functions of $m_{KK}$, and we have factored out the S-wave phase. We can then calculate the angular moments
\begin{eqnarray}
\sqrt{4\pi}\langle Y_0^0\rangle&=&\S_0^2+\P_0^2+\D_0^2+\P_{\pm 1}^2+\D_{\pm 1}^2\nonumber\\
\sqrt{4\pi}\langle Y_1^0\rangle&=&2\S_0\P_0\cos\Phi_{P_0}+\frac{4}{\sqrt{5}}\P_0\D_0\cos(\Phi_{P_0}-\Phi_{D_0}) \nonumber\\
&&+8\sqrt{\frac{3}{5}}\P_{\pm 1}\D_{\pm 1}\cos(\Phi_{P_{\pm 1}}-\Phi_{D_{\pm 1}}) \\
\sqrt{4\pi}\langle Y_2^0\rangle&=&\frac{2}{\sqrt{5}}\P_0^2+2\S_0\D_0\cos\Phi_{D_0}+\frac{2\sqrt{5}}{7}\D_0^2-\frac{1}{\sqrt{5}}\P_{\pm 1}^2+\frac{\sqrt{5}}{7}\D_{\pm 1}^2 \nonumber\\
\sqrt{4\pi}\langle Y_3^0\rangle&=&6\sqrt{\frac{3}{35}}\P_0\D_0\cos(\Phi_{P_0}-\Phi_{D_0})+\frac{6}{\sqrt{35}}\P_{\pm 1}\D_{\pm 1}\cos(\Phi_{P_{\pm 1}}-\Phi_{D_{\pm 1}})\nonumber\\
\sqrt{4\pi}\langle Y_4^0\rangle&=&\frac{6}{7}\D_0^2-\frac{4}{7}\D_{\pm 1}^2\nonumber.
\end{eqnarray}
The angular moments for $l>4$ vanish.
Figures \ref{fig:Sph_phi} and \ref{fig:Sph} show the distributions of the angular moments for the fit model around $\pm 30$ MeV of the $\phi(1020)$ mass peak and above the $\phi(1020)$, respectively. In general the interpretation of these moments is that  $\langle Y^0_0\rangle$ is the efficiency-corrected and background-subtracted event distribution, $\langle Y^0_1\rangle$  the sum of the interference between S-wave and P-wave, and P-wave and D-wave amplitudes, $\langle Y^0_2\rangle$  the sum of P-wave, D-wave and the interference of S-wave and D-wave amplitudes, $\langle Y^0_3\rangle$  the interference between P-wave and D-wave amplitudes, and $\langle Y^0_4\rangle$  the D-wave.
As discussed in Section~\ref{sec:Resonance_models}, the average of $\Bs$ and $\Bsb$ cancels the interference terms that involve $\P_0$ and $\D_{\pm}$. This causes the angular moments $\langle Y^0_1\rangle$ and $\langle Y^0_3\rangle$ to be zero when averaging over $\Bs$ and $\Bsb$ decays.
We observe that the fit results well describe the moment distributions, except for the $\langle Y^0_1\rangle$ and $\langle Y^0_4\rangle$ values  below 1.2 GeV. This may be the result of statistical fluctuations or imperfect modeling.

\subsection{Systematic uncertainties}
\label{sec:sys}
The sources of the systematic uncertainties on the results of the Dalitz plot analysis are summarized in Table~\ref{tab:sys_other}. The uncertainties due to the background parametrization are estimated by comparing the results from the best fit model with those  when the background shape parameters are obtained from  a fit to the lower sideband region only. The uncertainties in the efficiency are estimated by comparing the fit results when the efficiency parameters are changed by their statistical uncertainties and are added in quadrature. 
The effect on the fit fractions of changing the efficiency function is evaluated using a similar method to that used previously~\cite{LHCb:2012ae}. 
Briefly, we change the efficiency model
by increasing the minimum IP $\chi^2$ requirement from 9 to 12.5 on both of the kaon
candidates. This has the effect of increasing the $\chi^2$ of the fit to the angular 
distributions of $\Bsb \to \jpsi \phi$ data by 1 unit. The new efficiency function is then
applied to the data with the original minimum IP $\chi^2$ selection of 9, the likelihood is re-evaluated and the uncertainties are estimated by comparing the results with the best fit model. The largest variations among these two efficiency categories are included in the uncertainty.

\renewcommand{\arraystretch}{1.2}
\begin{table}[hbt]
\centering
\caption{Absolute systematic uncertainties on the fit results. }
\vspace{0.2cm}
\begin{tabular}{l|ccccc}
\hline
Item &    Efficiency &  Background & Fit model   & Total\\
\hline
$m_{f_2'(1525)}$~(MeV) & 1.2 & 0.4 &$^{+5.2}_{-1.5}$&$^{+5.3}_{-2.0}$  \\
$\Gamma_{f_2'(1525)}$~(MeV)&4.7& 0.5 &$^{+8.6}_{-1.8}$& $^{+9.8}_{-5.0}$ \\
\hline
\multicolumn{5}{c}{Fit fractions~(\%)}\\
\hline
$\phi(1020)~\lambda=0$  &0.8&0&$^{+0.06}_{-0.04}$& $\pm0.8$\\
$\phi(1020)~|\lambda|=1$  &1.3&0&$^{+0.4}_{-0.1}$&$\pm1.3$\\
$f_0(980)$  &1.7&0.4&$^{+2.1}_{-1.7}$&$^{+2.8}_{-2.5}$\\
$f_0(1370)$& 0.3&0.02&$^{+0.2}_{-1.2}$&$^{+0.3}_{-1.3}$\\
$f_2'(1525)~\lambda=0$ &1.5&0.2&$^{+1.8}_{-0.5}$&$^{+2.4}_{-1.6}$\\
$f_2'(1525)~|\lambda|=1$ &1.1&0.4&$^{+1.3}_{-0.8}$&$^{+1.8}_{-1.4}$\\
$f_2(1640)~|\lambda|=1$ &0.1&0.1&$^{+0.7}_{-0.9}$&$^{+0.7}_{-0.9}$\\
$\phi(1680)~|\lambda|=1$ &0.3&0.1&$^{+4.4}_{-0.1}$&$^{+4.4}_{-0.3}$\\
$f_2(1750)~\lambda=0$ &0.3&0.1&$^{+1.0}_{-0.5}$&$^{+1.0}_{-0.6}$\\
$f_2(1750)~|\lambda|=1$ &1.6&0.3&$^{+1.5}_{-2.9}$&$^{+2.2}_{-3.3}$\\
$f_2(1950)~\lambda=0$ &0.1&0.04&$^{+0.2}_{-0.4}$&$^{+0.2}_{-0.4}$\\
$f_2(1950)~|\lambda|=1$ &2.1&0.6&$^{+1.1}_{-3.1}$&$^{+2.5}_{-3.8}$\\
Non-resonant &1.7&0.4&$^{+0.9}_{-1.5}$&$^{+2.0}_{-2.2}$\\
\hline
S-wave within $\pm$12 MeV &0.02&0.02&$^{+0.2}_{-0.1}$&$^{+0.2}_{-0.1}$\\
of $\phi(1020)$ peak ~(\%) &&&&&\\
\hline
\multicolumn{5}{c}{Phases~(degrees)}\\
\hline
$f_0(980)$  &19&8&$^{+14}_{-~8}$&$^{+25}_{-22}$\\
$f_0(1370)$&6&1&$^{+6}_{-5}$&$\pm8$\\
$f_2(1640)~|\lambda|=1$ &7&11&$^{+~0}_{-42}$&$^{+13}_{-44}$\\
$\phi(1680)~|\lambda|=1$&1&1&$^{+260}_{-210}$&$^{+260}_{-210}$ \\
$f_2(1750)~\lambda=0$ &2&2&$^{+9}_{-9}$&$\pm 9$ \\
$f_2(1750)~|\lambda|=1$ &12&10&$^{+~2}_{-70}$&$^{+16}_{-70}$\\
$f_2(1950)~\lambda=0$&13&1&$^{+108}_{-~14}$&$^{+110}_{-~20}$ \\
$f_2(1950)~|\lambda|=1$ &77&26&$^{+130}_{-~~1}$&$^{+150}_{-~80}$\\
Non-resonant &18&7&$^{+12}_{-~5}$&$^{+23}_{-20}$\\
\hline         
\end{tabular}
\label{tab:sys_other}
\vspace*{8mm}
\end{table}
\renewcommand{\arraystretch}{1}

We estimate additional uncertainties by comparing the results when one more resonance is added to the best fit model. 
The uncertainties due to the line shape  of the contributing resonances with fixed mass and width parameters are estimated by varying them individually in the fit according to their combined statistical and systematic uncertainties added in quadrature.  We compare the results with the best fit and add them in quadrature to estimate the uncertainties due to the line shape. 

Another source of systematic uncertainty is the value we choose for $L_B$, the orbital angular momentum in the  $\Bsb$ decay. If $L_R$ equals zero then $L_B$ equals zero. If, however, $L_R$ is 1 then $L_B$ can either be 0 or 1, and if $L_R$ is 2, $L_B$ can be 1, 2 or  3. For our best fit we don't allow multiple values for $L_B$, but choose the lowest allowed value. 
To estimate the systematic uncertainties due to the choice of $L_B$, we repeat the fit changing the default value of $L_B$, in turn, to each higher allowed value and compare the fit results with the best fit. The differences are grouped into the fit model category, and we assign the largest variations as the systematic uncertainties. These later two categories often give in asymmetric uncertainties.

\section{Absolute branching fractions}
Branching fractions are measured from ratios of the decay rates of interest normalized to the well established decay mode $\Bm \to J/\psi \Km$.  This decay mode, in addition to having a well measured branching fraction, has the advantage of having two muons in the final state and hence the same triggers as the $\Bsb$ decay. However, we require knowledge of the $\Bsb/B^-$ production ratio. For this we assume isospin invariance and use the $\Bsb/\Bdb$ production ratio $f_s/f_d=0.256\pm0.020$, given in Ref.~\cite{fsfdhad,*Aaij:2011jp}.
The branching fractions are calculated using
\begin{equation}
\mathcal{B}(\Bsb \to J/\psi X)=\frac{N_{\Bsb}/\epsilon_{\Bsb}}{N_{\Bm}/\epsilon_{\Bm}}\times \mathcal{B}(\Bm \to J/\psi \Km) \times \frac{1}{f_s/f_d},\label{eq:br}
\end{equation}
where $X$ indicates a specific $K^+K^-$ state, $N$ represents the yield of the decay of interest, and $\epsilon$ corresponds to  the overall efficiency. We form an average of $\mathcal{B}(B^-\to \jpsi\Km)=(10.18\pm0.42)\times 10^{-4}$ using the recent \belle~\cite{Abe:2002rc} and \babar~\cite{Aubert:2004rz} measurements, corrected to take into account different rates of $\Bp\Bm$ and $\Bd\Bdb$ pair production from $\Upsilon(4S)$ using $\frac{\Gamma(\Bp\Bm)}{\Gamma(\Bd\Bdb)}=1.055\pm0.025$~\cite{PDG}.

The detection efficiency is obtained from simulation and is a product of the geometrical acceptance of the detector, the combined reconstruction and selection efficiency and the trigger efficiency. The efficiency also includes the efficiency of the Dalitz plot model for the case of $\Bsb \to J/\psi \Kp\Km$, where the best fit model is used.
The detection efficiencies and their various correction factors are given in Table~\ref{tab:MC_eff}. To ensure that the $p$ and $\pt$ distributions of the generated $B$ meson are correct we weight the $\Bsb$ simulations using $\Bsb\to J/\psi \phi(1020)$ data and the $\Bm$ simulations using $\Bm\to J/\psi \Km$ data.  Since the control channel has a different number of charged tracks than the decay channel,  we weight the simulations with the tracking efficiency ratio by comparing the data and simulations in bins of the track's $p$ and $\pt$. 
we further weight the $\Bsb \to \jpsi \KpKm$ simulation, using the PDG value of $\Bsb$ lifetime, $(1.497\pm0.015)\times 10^{-12}~\rm s$~\cite{PDG}, as input. 
  
\begin{table}[h!t!p!]
\centering
\caption{Detector efficiencies determined from simulation and the correction factors.}
\vspace{0.2cm}
\begin{tabular}{lcc}
\hline
Item & $J/\psi \Kp\Km$ & $J/\psi \Km$ \\
\hline
Detection efficiency (\%)& $1.061\pm0.004$  &$2.978\pm0.011$ \\
\hline
\multicolumn{3}{c}{Correction factors}\\
\hline
Tracking efficiency  & $0.999\pm 0.010$ & $1.003 \pm 0.010$\\
PID &$0.819\pm 0.008$& $0.974\pm 0.005$\\
$p$ and $\pt$ &$1.077\pm 0.005$&$1.053\pm 0.005$\\
$\Bsb$ lifetime & $0.993\pm0.015$ & - \\
\hline
Total efficiency (\%)&{$0.887\pm0.004\pm 0.018$}&{$3.065\pm0.012 \pm 0.038$}\\
\hline         
\end{tabular}
\label{tab:MC_eff}
\end{table}

The resulting branching fractions are
\begin{eqnarray*}
\mathcal{B}(\Bsb \to J/\psi \Kp\Km) ~~&=&  ~~\!(7.70\pm0.08\pm 0.39\pm 0.60)\times 10^{-4},\\
\mathcal{B}(\Bsb \to J/\psi \phi(1020))~ &=& (10.50\pm0.13\pm 0.64\pm 0.82)\times 10^{-4},\\
\mathcal{B}(\Bsb \to J/\psi f_2'(1525)) &=& ~~\!(2.61\pm 0.20^{+0.52}_{-0.46}\pm 0.20)\times 10^{-4},
\end{eqnarray*}
where the branching fractions ${\cal{B}}(\phi(1020)\to\Kp\Km)=(48.9\pm0.5)\%$ and ${\cal{B}}(f_2'(1525)\to\Kp\Km)=(44.4\pm 1.1)\%$ are used~\cite{PDG}.  Here the first uncertainty in each case is statistical, the second is systematic and the third reflects the uncertainty due to $f_s/f_d$. Note that these are the  time-integrated branching fractions. Results on the polarization fractions of $\Bsb\to\jpsi\phi(1020)$ from a time-dependent analysis will be forthcoming in a separate publication \cite{LHCb-PAPER-2013-001}.
The ratio of $\mathcal{B}(\Bsb \to J/\psi f_2'(1525))/\mathcal{B}(\Bsb \to J/\psi\phi(1020))$ is consistent with our previous result \cite{Aaij:2011ac}, D0 \cite{Abazov:2012dz}, and the \belle result~\cite{belleppt}. The current PDG value of $\mathcal{B}(\Bsb \to J/\psi \phi(1020))=(1.4\pm0.5)\times 10^{-3}$ is dominated by the \cdf measurement~\cite{Abe:1996kc}. Our measured value is in good agreement with this measurement and also the most recent yet unpublished values measured by \cdf~\cite{cdfnote:10795} and \belle~\cite{belleppt}.  The \belle collaboration has also recently reported the branching fraction of $\mathcal{B}(\Bsb \to J/\psi \Kp\Km)$~\cite{belleppt}, where $\Bsb\to J/\psi \phi(1020)$ is excluded. 




       
\begin{table}[b]
\centering
\caption{Relative systematic uncertainties on branching fractions~(\%).}
\vspace{0.2cm}
\begin{tabular}{lcccc}
\hline
Item & $J/\psi \Kp\Km$ & $J/\psi \phi(1020)$ & $J/\psi f_2'(1525)$ \\
\hline 
Tracking efficiency & 1.0&1.0 &1.0\\
Material and physical effects &2.0 & 2.0& 2.0\\
PID efficiency & 1.0 & 1.0  & 1.0 \\
$\Bsb$ $p$ and $\pt$ distributions & 0.5 &0.5 &0.5\\
$\Bm$ $p$ and $\pt$ distributions & 0.5 &0.5 &0.5\\
$\Bsb$ lifetime & 1.5 & 1.5 & 1.5 \\
Efficiency function & 0.02 & 0.02&0.02\\
$\mathcal{B}(\phi(1020)\to \Kp\Km)$&-&1.0&-\\
$\mathcal{B}(f_2'(1525)\to \Kp\Km)$&-&-&2.5\\
$\mathcal{B}(\Bm\to J/\psi\Km)$&4.1&4.1&4.1\\
\hline
\multicolumn{4}{c}{Contributions from Dalitz analysis}\\
\hline
Efficiency &-&3.1&12.3\\
Background &-&0&1.3\\
Fit model &-&$^{+0.7}_{-0.2}$&$^{+14.6}_{-11.2}$\\[.2ex]
\hline 
\multicolumn{4}{c}{}\\[-2.4ex]
Sum in quadrature of items above &5.0&$^{+6.1}_{-6.0}$&$^{+20.0}_{-17.7}$\\[.2ex]
\hline
$f_s/f_d$ &$7.8$&$7.8$&$7.8$\\
\hline        
\end{tabular}
\label{tab:sys_br}
\end{table}

The systematic uncertainty on the branching fraction has several contributions listed in Table~\ref{tab:sys_br}. Since the branching fractions are measured with respect to $\Bm\to J/\psi\Km$ which has a different number of charged tracks than the decays of interest, a 1\% systematic uncertainty is assigned due to differences in the tracking performance between data and simulation. Another 2\% uncertainty is assigned for the additional kaon which is due to decay in flight, large multiple scatterings and hadronic interactions along the track. 
Using the PDG value for the $\Bsb$ lifetime~\cite{PDG} as input gives rise to an additional 1.5\% systematic uncertainty. 
 Small uncertainties are introduced if the simulation does not have the correct $\B$ meson kinematic distributions. We are relatively insensitive to any of these differences in the $\B$ meson $p$ and $\pt$ distributions since we are measuring the relative rates. By varying the $p$ and $\pt$ distributions we see at most a change of 0.5\%. There is a 1\% systematic uncertainty assigned for the relative particle identification efficiencies. 
An uncertainty of 0.02\% is included due to the change of the efficiency function Eq.~(\ref{eq:eff}). Three additional uncertainties are considered in the  branching fractions of $\mathcal{B}(\Bsb\to J/\psi \phi(1020))$ and $\mathcal{B}(\Bsb\to J/\psi f_2'(1525))$ as these are measured from the fit fractions of the Dalitz plot analysis.  The total systematic uncertainty is obtained by adding each source of systematic uncertainty in quadrature as they are uncorrelated.

\section{Conclusions}
We have determined the final state composition of the $\Bsb \to J/\psi \Kp\Km$ decay channel using a modified Dalitz plot analysis where we include the decay angle of the $J/\psi$. The largest contribution is  the $\phi(1020)$ resonance, along with other S-, P- and D-wave $\KpKm$ states,  and a non-resonant $\KpKm$ contribution. All of the components are listed in Table~\ref{tab:fitfraction}. The mass and width of the $f_2'(1525)$ resonance are measured as
\begin{eqnarray*}
 m_{f_2'(1525)} &=& 1522.2 \pm 2.8^{+5.3}_{-2.0}~\rm MeV, \\
\Gamma_{f_2'(1525)} &=& 84 \pm 6^{+10}_{-~5} ~ \rm MeV.
\end{eqnarray*}
We also observe a significant S-wave component that is present over the entire $K^+K^-$ mass region. Within $\pm$12~MeV of the 
$\phi(1020)$ mass it is  $(1.1\pm 0.1 ^{+0.2}_{-0.1})\%$ of the yield, and can affect precision \CP violation measurements~\cite{Stone:2008ak}. Finally we determine the absolute branching fractions
\begin{eqnarray*}
\mathcal{B}(\Bsb \to J/\psi \Kp\Km) ~~&=&  ~~\!(7.70\pm0.08\pm 0.39\pm 0.60)\times 10^{-4},\\
\mathcal{B}(\Bsb \to J/\psi \phi(1020))~ &=& (10.50\pm0.13\pm 0.64\pm 0.82)\times 10^{-4},\\
\mathcal{B}(\Bsb \to J/\psi f_2'(1525)) &=& ~~\!(2.61\pm 0.20^{+0.52}_{-0.46}\pm 0.20)\times 10^{-4},
\end{eqnarray*}
where the first uncertainty in each case is statistical, the second is systematic and the third due to $f_s/f_d$. These results provide a good understanding of the $\jpsi K^+K^-$ final state in $\Bsb$ decays over the entire kinematically allowed region.  The  $J/\psi f_2'(1525)$ results supersede those of Ref.~\cite{Aaij:2011ac}. This decay mode offers the opportunity for additional measurements of \CP violation \cite{Sharma:2005ji,*Zhang:2012zk}.
\newpage
\section*{Acknowledgements}

\noindent We express our gratitude to our colleagues in the CERN
accelerator departments for the excellent performance of the LHC. We
thank the technical and administrative staff at the LHCb
institutes. We acknowledge support from CERN and from the national
agencies: CAPES, CNPq, FAPERJ and FINEP (Brazil); NSFC (China);
CNRS/IN2P3 and Region Auvergne (France); BMBF, DFG, HGF and MPG
(Germany); SFI (Ireland); INFN (Italy); FOM and NWO (The Netherlands);
SCSR (Poland); ANCS/IFA (Romania); MinES, Rosatom, RFBR and NRC
``Kurchatov Institute'' (Russia); MinECo, XuntaGal and GENCAT (Spain);
SNSF and SER (Switzerland); NAS Ukraine (Ukraine); STFC (United
Kingdom); NSF (USA). We also acknowledge the support received from the
ERC under FP7. The Tier1 computing centres are supported by IN2P3
(France), KIT and BMBF (Germany), INFN (Italy), NWO and SURF (The
Netherlands), PIC (Spain), GridPP (United Kingdom). We are thankful
for the computing resources put at our disposal by Yandex LLC
(Russia), as well as to the communities behind the multiple open
source software packages that we depend on.

\newpage
\ifx\mcitethebibliography\mciteundefinedmacro
\PackageError{LHCb.bst}{mciteplus.sty has not been loaded}
{This bibstyle requires the use of the mciteplus package.}\fi
\providecommand{\href}[2]{#2}

\end{document}